\documentclass[pra,twocolumn,footinbib,showpacs,letterpaper]{revtex4}
\usepackage{graphicx}
\usepackage{hyperref}
\usepackage{rotating}
\hypersetup{colorlinks,citecolor=blue}

\begin{document}

\setlength{\pdfpageheight}{\paperheight}
\setlength{\pdfpagewidth}{\paperwidth}

\hyphenation{Gauss-ian}

\pacs{37.10.Vz, 37.10.Gh, 03.75.Be}
\title{Dynamics of Cold Atoms Crossing a One-Way Barrier}
\author{Jeremy~J.~Thorn, Elizabeth~A.~Schoene, Tao~Li, and Daniel~A.~Steck}
\affiliation{Oregon Center for Optics and Department of Physics, 1274 University
of Oregon, Eugene, Oregon 97403-1274, USA}

\begin{abstract}
We implemented an optical one-way potential barrier that allows ultracold
$^{87}\mathrm{Rb}$ atoms to transmit through when incident on one side of
the barrier but reflect from the other.
This asymmetric barrier is a realization of Maxwell's demon, which can be
employed to produce phase-space compression and has implications for
cooling atoms and molecules not amenable to standard laser-cooling
techniques.
The barrier comprises two focused Gaussian laser beams that intersect the
focus of a far-off-resonant single-beam optical dipole trap that holds the
atoms.
The main barrier beam presents a state-dependent potential to incident
atoms, while the repumping barrier beam optically pumps atoms to a trapped
state.
We investigated the robustness of the barrier asymmetry to changes in the
barrier-beam separation, the initial atomic potential energy, the intensity
of the second beam, and the detuning of the first beam.
We performed simulations of the atomic dynamics in the presence of the
barrier, showing that the initial three-dimensional momentum distribution
plays a significant role, and that light-assisted collisions are likely the
dominant loss mechanism.
We also carefully examined the relationship to Maxwell's demon and
explicitly accounted for the apparent decrease in entropy for our
particular system.
\end{abstract}

\maketitle

\section{\label{sec:introduction}Introduction}

The transport dynamics induced by \emph{asymmetries} of systems is an
important paradigm in physics, dating back to Feynman's thought-experiment
of a mechanical ratchet at finite temperature \cite{feynman1963}.
The point of Feynman's analysis was to demonstrate that in thermal
equilibrium, the ratchet cannot be used to ``rectify'' thermal fluctuations
to do useful work, because Brownian motion of the ratchet mechanism itself
causes it to intermittently fail in such a way as to prevent steady-state
motion against a constant force.
However, in \emph{nonequilibrium} situations---such as a simple
temperature difference---the asymmetry of a system \emph{can} induce
steady-state transport, and  of course do work.
Recently, the study of more general ``ratchets,'' where thermal
fluctuations or time-dependent forces couple with periodic asymmetric
potentials to produce steady-state directed motion \cite{haenggi1996}, has
become of broad interest.
Ratchet systems in the form of molecular and Brownian motors
\cite{julicher1997,linke2005,leigh2007} are of particular interest in
understanding how controlled motion is effected in spite of thermal
fluctuations, especially at the nanoscale, where such fluctuations tend to
be large.  Ratchets have also recently been studied in the context of the
laser-cooling of atoms in asymmetric optical-lattice potentials
\cite{mennerat-robilliard1999}.

A related problem of asymmetric transport occurs for particles in the
presence of asymmetric barriers.
For example, asymmetric diffusion occurs in transport across membranes.
A recent experiment demonstrated that asymmetric transport can be caused by
an asymmetry in the shape of the membrane pores, provided that a
sufficiently wide range of particle sizes is present \cite{shaw2007}.
In this case, larger particles can clog the pores (but only from one
side), preventing smaller particles from diffusing through the membrane in
one direction.
In atom optics, it is possible to realize similar ``one-way barriers'' or
``atom diodes,'' as proposed independently by Raizen
\textit{et~al.} \cite{raizen2005} and Ruschhaupt and Muga
\cite{ruschhaupt2004} in slightly different contexts (with a number of
subsequent refinements in the general design
\cite{ruschhaupt2006,ruschhaupt2006b,ruschhaupt2007}).
These optical one-way barriers for cold atoms are \emph{asymmetric} optical
potentials in the sense that atoms experience a different potential
depending on the direction of incidence: atoms incident from one side
reflect from the barrier, while atoms  incident from the other side
transmit through it.
Unlike the above membrane, however, this one-way action is entirely a
single-particle effect, relying on optical potentials that change depending
on the internal state of the atom.
Note that in a closely related proposal, a combination of coherent and
incoherent processes conspires to allow transport of particles from one
reservoir to another, but not in the reverse direction \cite{kim2005}.

A common theme among all the systems mentioned above is that
\emph{dissipation} is required to produce asymmetric transport.
Ratchets, for example, typically involve overdamped motion and thus heavy
dissipation.
However, an interesting question is, how \emph{little} dissipation is
required to produce a significant asymmetry in the transport?
The most straightforward answer for the optical one-way barrier here is
that one photon must be spontaneously scattered per particle, since the
action of the one-way barrier depends on an irreversible change in the
internal state of the atom.
(Presumably, though, one can construct more complicated schemes that
require even less dissipation; we will show below that our one-way barrier
dissipates more entropy than is required by the second law of
thermodynamics.)
This question is particularly relevant in the atom-optical case, since one
of the main motivations in realizing one-way barriers for atoms is in
developing new laser-cooling methods for atoms
\cite{raizen2005,dudarev2005,ruschhaupt2006c,price2007,ruschhaupt2008}.
The idea here is that the one-way barrier can force the accumulation of
atoms into a volume much smaller than the original container.
The phase-space density of the atomic vapor increases if the effect of the
volume compression outweighs any heating effects of the barrier.
(Cooling can then be straightforwardly effected via adiabatic expansion;
the most important step in terms of controlling an atomic vapor is
reducing the phase-space volume.)
Phase-space compression with one-way barriers has recently been
demonstrated with rubidium atoms \cite{price2008,thorn2008}.
The best increase in phase-space density achieved so far is a factor of
$350$ above the initial conditions of a vapor in a magnetic trap
\cite{bannerman2009}.
Though standard laser-cooling techniques such as Doppler cooling are now
well established \cite{metcalf1999}, they generally rely on the existence
of a cycling optical transition: atoms falling into ``dark states''
decouple from the cooling lasers.
Since the irreversible action of a one-way barrier relies in principle only
on the scattering of a single photon, laser-cooling techniques based on
one-way barriers may help circumvent problems with dark states and enable
laser cooling of molecules \cite{narevicius2009} and the many atoms for
which standard laser-cooling methods are ineffective.

Another, more fundamental, motivation for studying the one-way barrier
comes from its connection to Maxwell's demon
\cite{maxwell1871,bennett1987,scully2007,ruschhaupt2006c}.
In this thought-experiment, the demon manipulates a trapdoor in a wall
dividing a container of gas.
The demon could use the trapdoor as a one-way barrier to reduce the space
occupied by the atoms without performing any work, in an apparent violation
of the second law of thermodynamics.
(Maxwell's original demon used the trapdoor to separate hot and cold atoms
to achieve a temperature gradient, but the reduction of entropy is the
essence of the thought-experiment.)
One can think about the resolution to this ``paradox'' in a number of ways
\cite{bennett1987,scully2007}, but the key issue is that the demon must
perform measurements on the system, gaining the information needed to
perform feedback via the trapdoor.
The entropy decrease of the atoms is balanced by an \emph{increase} in the
entropy of the demon's memory due to the accumulated information.
This cannot continue indefinitely in a cyclic process unless the demon's
memory is reset, or ``erased.''
The erasure requires transferring entropy to the environment, in accordance
with the second law.
Our experiment effectively consists of only a single thermodynamic cycle,
and so a cyclic erasure is not strictly necessary.
The spontaneous scattering of photons from the barrier laser light in our
experiment both acts as an effective position measurement on each atom and
carries away sufficient entropy to compensate for entropy lost in the
reduction in phase-space volume.

This work builds on our previous study of an all-optical realization of a
one-way barrier for rubidium atoms \cite{thorn2008}.
Here we perform a detailed study of the dynamics of cold atoms crossing the
one-way barrier.
We study how various aspects of the barrier configuration contribute to its
operation, and we discuss the key elements of optimizing the robustness and
performance of the barrier.
We also compare our data to simulations that treat the center-of-mass
motion of the atoms classically, and obtain good agreement.
Finally, we examine in some detail the entropic aspects of phase-space
compression with the one-way barrier.

\section{\label{sec:procedures}Procedures}

\begin{figure}[tbp]
	\begin{center}
		\includegraphics[width=\columnwidth]{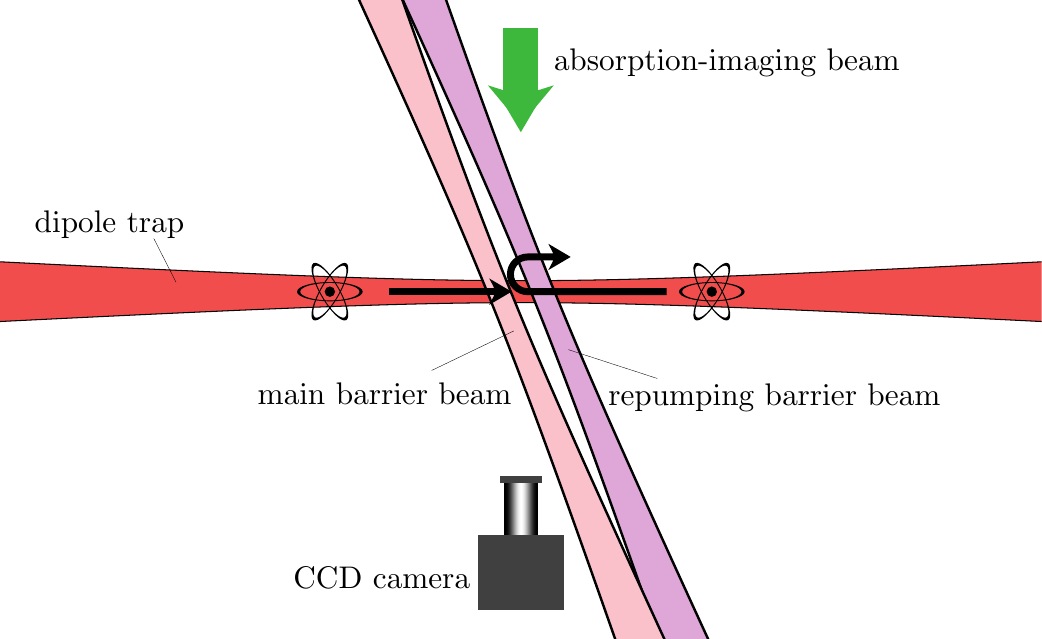}
	\end{center}
	\vspace{-5mm}
	\caption{%
		(Color online)
		Schematic diagram of the optical setup, showing the
 		dipole trap, barrier beams, and imaging system.
	\label{fig:opticalsetup}}
\end{figure}

\begin{figure}[tbp]
	\begin{center}
		\includegraphics[scale=1]{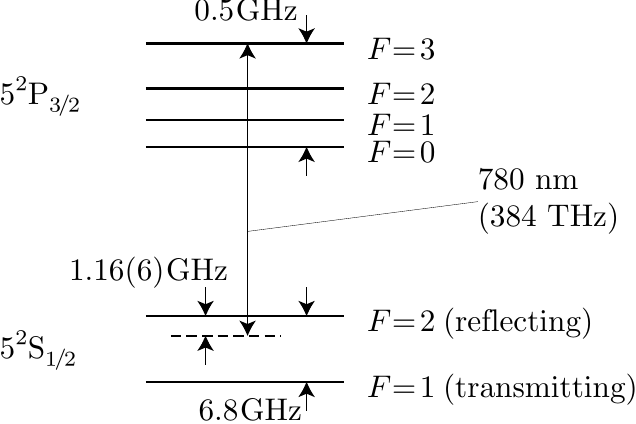}
	\end{center}
	\vspace{-5mm}
	\caption{%
		Relevant atomic levels of $^{87}\mathrm{Rb}$ on the $\mathrm{D}_2$
		cooling transition (not to scale), shown with the off-resonant
		coupling of the main barrier beam.
		Atoms with $F{=}1$ see this field as red-detuned, while $F{=}2$ atoms
		see a blue detuning.
		The repumping barrier beam is resonant with the $F{=}1 \to F'{=}2$ MOT
		repumping transition.
	\label{fig:hfs}}
\end{figure}

\begin{figure}[tbp]
	\begin{center}
		\includegraphics[width=\columnwidth]{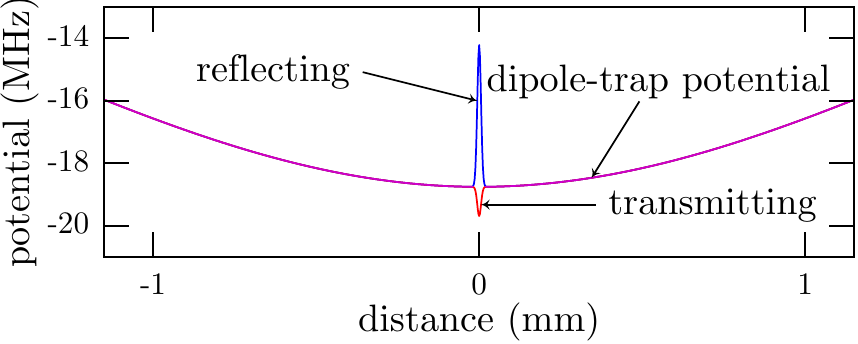}
	\end{center}
	\vspace{-5mm}
	\caption{%
		(Color online)
		Optical potentials seen by atoms along the axis of the dipole trap in
		our experiment.
		The overall trapping potential is from the main dipole-trap beam, while
		the narrow state-dependent feature results from the main barrier beam.
		The main barrier beam produces a repulsive potential for atoms in the
		reflecting ($F{=}2$) ground state, and an attractive potential for
		atoms in the transmitting ($F{=}1$) ground state.
	\label{fig:potential}}
\end{figure}

The optical one-way barrier is the principal feature of our experimental
setup (Fig.~\ref{fig:opticalsetup}), comprising two focused Gaussian laser
beams.
The one-way barrier scheme exploits the hyperfine ground-state structure of
$^{87}\mathrm{Rb}$ to create an asymmetry that allows atoms to transmit
through the barrier when traveling in one direction, but reflects them when
traveling in the other. 
The main barrier beam is tuned between the $F{=}1 \to F'$ and $F{=}2 \to
F'$ $^{87}\mathrm{Rb}$ hyperfine transitions, as shown in
Fig.~\ref{fig:hfs}.
Thus the detuning $\Delta:=\omega-\omega_0$ of the barrier-beam frequency
$\omega$ with respect to the atomic resonance $\omega_0$ has opposite signs
for atoms in the two hyperfine ground levels.
This results in the main barrier beam presenting an \emph{attractive}
potential to atoms in the $F{=}1$ ground state and a \emph{repulsive}
potential to atoms in the $F{=}2$ ground state, because the optical dipole
potential is inversely proportional to the detuning $\Delta$.
These two potentials are plotted in Fig.~\ref{fig:potential}.
To create the asymmetry, we introduce a second beam, the repumping
barrier beam, that is resonant with the $F{=}1 \to F'{=}2$ repumping
transition, displaced from the main barrier beam as shown in
Fig.~\ref{fig:opticalsetup}.
If all atoms start in the transmitting ($F{=}1$) ground state, they can
pass through the main barrier beam until they pass through the repumping
barrier beam, at which point they will be pumped to the reflecting
($F{=}2$) ground state.
The atoms will then see the main barrier beam as a potential barrier and
will remain trapped on that side of the one-way barrier.
We choose to have the repumping barrier beam on the right-hand side of the
main barrier beam, which makes that side the reflecting side of the
barrier, and the left-hand side the transmitting side.

We initially cool and trap the $^{87}\mathrm{Rb}$ atoms in a standard
six-beam magneto-optic trap (MOT) \cite{metcalf1999}, loaded from a cold
atomic beam produced by a pyramid MOT \cite{lee1996}.
After a secondary polarization-gradient cooling stage with reduced
intensity and increased detuning of the trapping beam, we have about $2
\times 10^5$ atoms at about $30 \ \mathrm{\mu K}$ in an ultra-high vacuum
of $\lesssim 10^{-10} \ \mathrm{torr}$.
The position of the MOT can be shifted several millimeters in each
coordinate using magnetic bias fields.

After cooling we load the atoms into a far-detuned optical dipole trap
produced by a $1090(5) \ \mathrm{nm}$ Yb:fiber laser.
The laser emits a collimated, multiple-longitudinal-mode, unpolarized,
nearly Gaussian beam with a $1.9(1) \ \mathrm{mm}$ $1/e^2$ beam radius.
We operate this laser so that the total power inside the vacuum chamber is
$9.3(5) \ \mathrm{W}$.
We focus the beam with a single $200 \ \mathrm{mm}$ focal length
plano-convex lens, producing a $30.9(5) \ \mathrm{\mu m}$ waist ($1/e^2$
intensity radius) and a $2.8 \ \mathrm{mm}$ Rayleigh length.
For $^{87}\mathrm{Rb}$ atoms in either hyperfine ground state, this beam
yields a nearly conservative potential well with a maximum potential depth
of $k_{_\mathrm{B}} \times 0.9 \ \mathrm{mK}$.
This dipole trap has longitudinal and transverse harmonic frequencies of
$24 \ \mathrm{Hz}$ and $3.0 \ \mathrm{kHz}$, respectively (near the trap
center), a $1/e$ lifetime of $20 \ \mathrm{s}$, and a maximum scattering
rate of only $3 \ \mathrm{s^{-1}}$.
Atoms are typically loaded in the anharmonic region of the trap, so that
the atomic motion dephases and has a different period ($50 \ \mathrm{ms}$)
than the harmonic frequency suggests (angular momentum adds to this effect;
see Appendix \ref{app:ang_mom}).

The two one-way barrier beams are nearly parallel asymmetric Gaussian beams
with a variable separation.
Their foci nearly coincide with the focus of the dipole-trap beam,
intersecting it at about $12(3)^\circ$ from the perpendicular to the beam
axis (as in Fig.~\ref{fig:opticalsetup}).
The main [repumping] barrier beam has a waist of $11.5(5) \ \mathrm{\mu m}$
[$13(2) \ \mathrm{\mu m}$] along the dipole-trap axis and $80(7) \
\mathrm{\mu m}$ [$60(7) \ \mathrm{\mu m}$] perpendicular to the dipole-trap
axis.
We control the power of the repumping barrier beam with an acousto-optic
modulator (AOM).

We measure the separation of the two beams on a beam profiler that has
a resolution of $5.6 \ \mathrm{\mu m/pixel}$.
We take pictures of each beam profile at many locations along the beam axis
near the focus, thus making it possible to accurately locate the axial
position of the focus despite the fact the beam waists are on the same
order of magnitude as the camera resolution.
Using this setup, we determine the separation of the beams (on the order of
ten microns) with an error on the order of a micron.

We image the atoms by illuminating them with a $45 \ \mathrm{\mu s}$ pulse
of light resonant with the $F{=}2 \to F'{=}3$ MOT trapping transition.
This absorption-imaging beam is nearly perpendicular to the dipole-trap
beam, and is detected by a charge-coupled-device (CCD) camera, which images
the shadow left by the atoms that scatter light out of the beam.
We image atoms in the $F{=}1$ ground state (which do not scatter the MOT
trapping light) by turning the MOT repumping beams on slightly before and
throughout the imaging pulse, ensuring that all atoms are in the $F{=}2$
ground state.
The short image pulse is ideal for getting accurate spatial imaging, since
the atoms cannot move much during that time.
To reduce systematic errors due to interference fringes in the images, we
subtract background offsets (as computed from the edge regions of the
images) on a per-column basis and then integrate each column to form
distributions (such as the ones shown in Fig.~\ref{fig:waterfall}).
The spatial resolution is $24.4 \ \mathrm{\mu m}$ as set by the CCD pixel
spacing, but the distributions are smoothed slightly for visual clarity.
The total number of trapped atoms drifted slowly over time, so we often
rescaled atom distributions in order to aid comparison of different data
sets.

Our ``canonical'' data are taken as follows.
We load the dipole trap with the MOT centered on the dipole-beam axis
$0.95(5) \ \mathrm{mm}$ away from the dipole-trap focus for $5 \
\mathrm{ms}$, trapping about $3 \times 10^4$ atoms (measured by imaging the
resonance fluorescence from the MOT light on a CCD camera) at $\sim\! 100 \
\mathrm{\mu K}$, with a peak one-dimensional atom density of around $4
\times 10^4 \ \mathrm{atoms/mm}$ (on the order of $10^7 \
\mathrm{atoms/mm^3}$).
Longer load times trap more atoms, but the atoms spread throughout the trap
during loading.
We used a relatively short loading time to keep the atoms localized.
To allow atoms from the MOT that are not trapped in the dipole trap to fall
away, we extinguish the MOT beams and wait approximately $20 \ \mathrm{ms}$
(about half an oscillation period) after loading atoms into the dipole beam
but before turning the barrier beams on.
During this half-period delay, atoms spread in the dipole trap, reducing
their apparent mean displacement from the dipole-trap focus.
For example, the center of the atomic distribution for the normal $0.95(5)
\ \mathrm{mm}$ mean starting offset of the atoms is only $0.58(8) \
\mathrm{mm}$ away from the dipole-trap focus after the half-period delay.
Henceforth, we will quote the offset of the center of the atomic
distribution after this delay.

During the half-period delay after loading atoms into the dipole trap, we
optically pump the loaded atoms to the $F{=}1$ ground state by extinguishing
the MOT repumping beams and then leaving the MOT trapping beams
(red-detuned by $70 \ \mathrm{MHz}$) on for another $7 \ \mathrm{ms}$.
We verified that this pumps virtually all the atoms to the $F{=}1$ ground
state by using absorption imaging without the MOT repumping beams.

After the half-period delay, we turn on the barrier beams (we designate
this time as $t=0$), allow the system to evolve for some time ranging from
a few milliseconds to as long as a few seconds, and then image the atoms.
We load atoms to either side of the barrier beams, and label data as either
``transmitting side'' or ``reflecting side,'' based on which side of the
barrier beams the atoms were on at $t=0$ (when the barrier beams are turned
on).
Because we wait a half-period between loading the atoms and $t=0$, atoms
are actually loaded on the \emph{opposite} side of the barrier beams with
respect to their ``starting'' location.

The main barrier beam was normally $34(1) \ \mathrm{\mu m}$ to the left (as
seen by the camera) of the repumping barrier beam, making the left-hand
(right-hand) side of the barrier beam the transmitting (reflecting) side.
We normally ran the main barrier beam with $40(4) \ \mathrm{\mu W}$ of
power (inside the vacuum chamber), resonant with the $^{85}\mathrm{Rb}$
$F{=}3 \to F'{=}3,4$ crossover dip in the rubidium saturated-absorption
spectrum, which is $1.05(5) \ \mathrm{GHz}$ blue of the $^{87}\mathrm{Rb}$
MOT trapping transition.
We normally ran the repumping barrier beam at $0.36(4) \ \mathrm{\mu W}$
(inside the vacuum chamber), and always stabilized it to the
$^{87}\mathrm{Rb}$ $F{=}1 \to F'{=}2$ MOT repumping transition.

We use two different conventions when reporting laser detunings.
When we wish to give the absolute frequency of a laser, we report the
detuning as the difference between the laser frequency and the $F{=}2 \to
F'{=}3$ $^{87}\mathrm{Rb}$ MOT transition.
However, for computing optical potentials and scattering rates, we need to
account for the presence of four separate excited states, and the hyperfine
splitting between these states is not negligible.
For these purposes, we compute and report an effective detuning.
The dipole potential for each two-level transition is proportional to the
squared dipole matrix element for the transition, and inversely
proportional to the detuning.
In order to keep a two-level-atom approximation, we average over the
different excited states, using the correct dipole matrix elements and
energy shifts, to obtain an effective two-level detuning.
For example, in our canonical setup, our main barrier beam is tuned
$1.05(5) \ \mathrm{GHz}$ to the blue of the $F{=}2 \to F'{=}3$
$^{87}\mathrm{Rb}$ MOT transition.
When we compute an optical potential using all four excited states in the
$\mathrm{D}_2$ manifold to obtain an effective two-level-atom detuning that
gets the same answer, we find $1.12 \ \mathrm{GHz}$ for atoms in the
$F{=}2$ ground state, and $-5.41 \ \mathrm{GHz}$ for atoms in the $F{=}1$
ground state.
For computing scattering rates, we need to average squared dipole matrix
elements divided by the square of the detuning.
For the detunings we used, effective two-level detunings for both optical
potentials and scattering rates differed by less than the accuracies quoted
here.

One common variation from our canonical setup was the ``$F{=}2$ setup.''
For these data, during the half-period while we waited for MOT atoms to
fall away, we chose to optically pump the atoms into the $F{=}2$ ground state
by flashing the MOT repumping beam without the MOT trapping beam.
Unless otherwise specified, we optically pumped atoms into the $F{=}1$
ground state.

Another variation we used was the ``filled trap.''
For these experiments we loaded the atoms to the left-hand side of the
barrier beams for $110 \ \mathrm{ms}$ and waited $200 \ \mathrm{ms}$ before
turning the barrier beams on, loading about $9 \times 10^4$ atoms at
$\sim\! 100 \ \mathrm{\mu K}$.
These times are several times the longitudinal trap period, and so resulted
in a fairly even distribution of atoms throughout the dipole trap.
We used the same method described above to ensure all atoms were pumped
into the $F{=}1$ ground state for these experiments.

Unless otherwise noted, the main barrier beam was linearly polarized
perpendicular to the dipole-trap beam axis, and the repumping barrier beam
was linearly polarized parallel to the dipole-trap beam axis.
We found that beam polarization was not a critical factor in the barrier
operation.

\section{\label{sec:main_results}Main Results}

\begin{turnpage}\begin{figure*}[p]
	\begin{center}
		\includegraphics[width=\textheight]{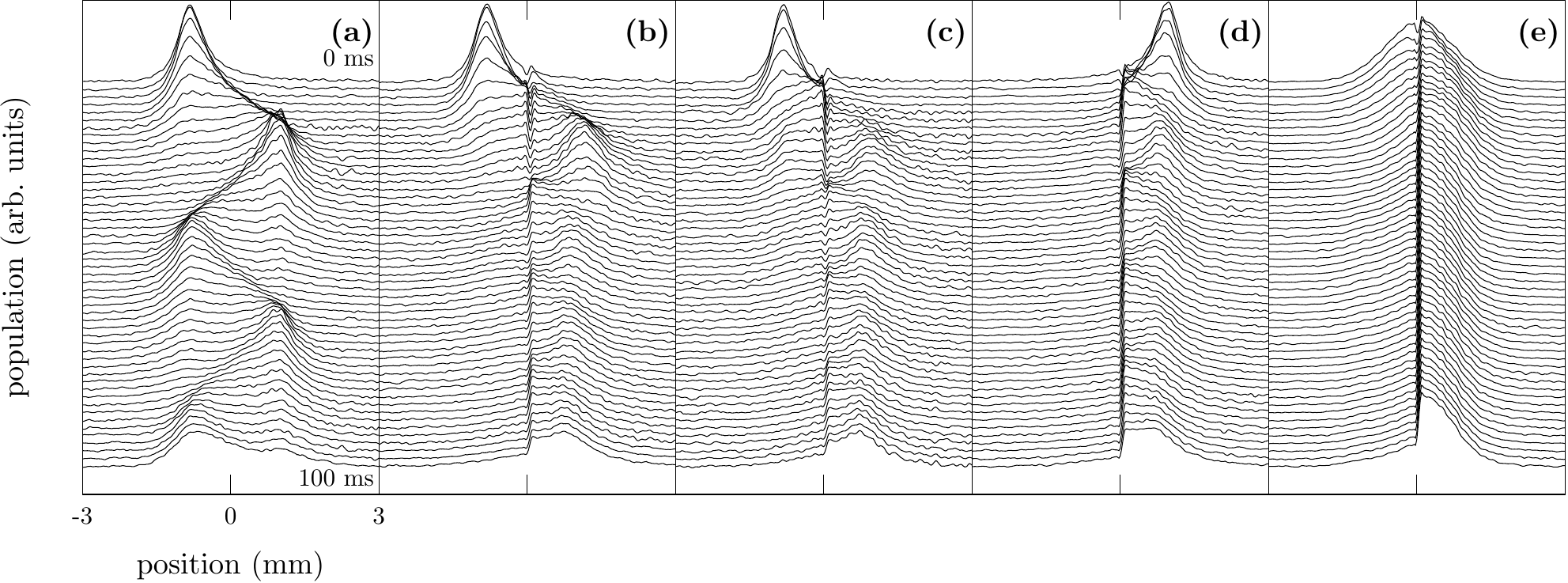}
	\end{center}
	\vspace{-5mm}
	\caption{%
		Density of atoms along the dipole trap responding to the one-way
		barrier.
		The dipole-trap focus and barrier beams are located at the horizontal
		origins.
		Each curve represents an average of $78$ [$38$ for column (e)]
		repetitions of the experiment.
		Column (a): atoms are initially in the $F{=}1$ (transmitting) state and
		to the left of the trap center, with no barrier present.
		Column (b): atoms are initially in the $F{=}1$ (transmitting) state on
		the left (transmitting) side of the barrier.
		Column (c): atoms are initially in the $F{=}2$ (reflecting) state on
		the left (transmitting) side of the barrier.
		Column (d): atoms are initially in the $F{=}1$ (transmitting) state on
		the right (reflecting) side of the barrier.
		Column (e): atoms are initially loaded relatively uniformly throughout
		the dipole trap (a ``filled trap'') in the $F{=}1$ (transmitting)
		state.
	\label{fig:waterfall}}
\end{figure*}\end{turnpage}

\begin{figure}[tbp]
	\begin{center}
		\includegraphics[width=\columnwidth]{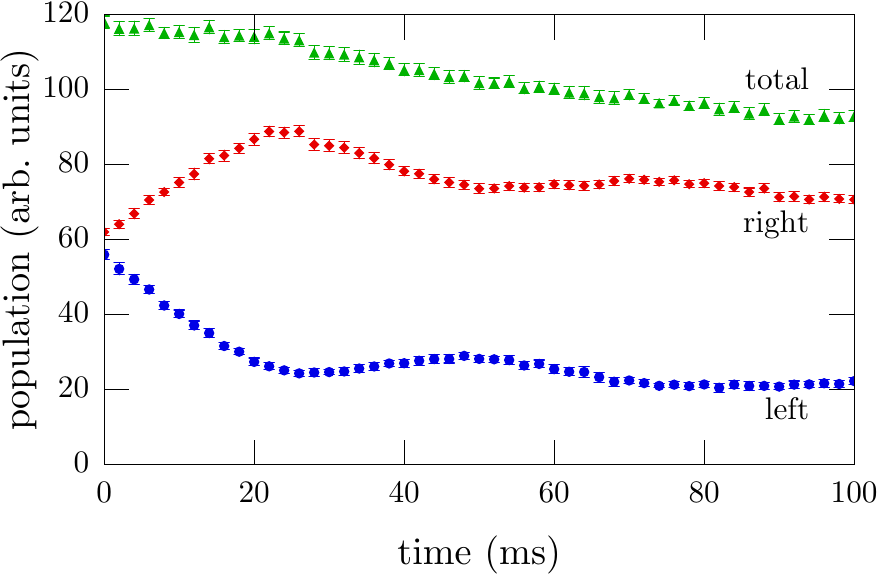}
	\end{center}
	\vspace{-5mm}
	\caption{%
		(Color online)
		Populations on the left- and right-hand sides of the one-way barrier
		for a ``filled-trap'' experiment where atoms were initially distributed
		throughout the dipole trap.
		Error bars indicate statistical error from $38$ repetitions.
	\label{fig:compression}}
\end{figure}

The main results for the one-way barrier are presented in
Fig.~\ref{fig:waterfall}, which shows the atomic evolution in response to
the one-way barrier.
To elucidate the effects of the one-way barrier,
Fig.~\ref{fig:waterfall}(a) shows the evolution of the atoms in the
dipole trap in the absence of the barrier.
As expected, the atoms oscillate back and forth about the trap center with
some breakup of the atomic cloud evident at later times due to the
anharmonicity of the trapping potential and the spread in angular momentum
of atoms orbiting the dipole-trap axis.

Figure \ref{fig:waterfall}(b) shows the dynamics of the atoms in the presence
of the one-way barrier, which is located at the origin.
The repumping barrier beam is positioned to the right of the main barrier
beam in the columns so that the atoms, initially released from the
left-hand side of the barrier, will interact first with the main barrier
beam.
The repumping barrier beam defines the reflecting side of the barrier,
which makes the left side the transmitting side.
The atoms, which are initially prepared in the $F{=}1$ transmitting state,
pass through the main barrier beam to the right-hand side of the barrier.
There the atoms interact with the repumping barrier beam, which optically
pumps them to the $F{=}2$ reflecting state.
Upon their return, the atoms are effectively trapped on the right-hand side
of the barrier as they repeatedly reflect from the barrier.

When the atoms start on the reflecting side of the barrier, as in
Fig.~\ref{fig:waterfall}(d), they also remain trapped to the right of the
barrier.
This holds true regardless of the initial state of the atoms; atoms
starting in the $F{=}2$ reflecting state are expected to reflect from the
main barrier beam, while atoms starting in the $F{=}1$ transmitting state
will first encounter the repumping barrier beam, which will optically pump
them to the $F{=}2$ reflecting state before they reach the main barrier
beam.
The results presented in Fig.~\ref{fig:waterfall}(d) show the evolution for
atoms initially in the $F{=}1$ transmitting state, and we observe similarly
good reflections for atoms starting in the $F{=}2$ reflecting state.

Figure \ref{fig:waterfall}(c) reveals what happens when atoms start on the
left-hand (transmitting) side of the barrier in the $F{=}2$ (reflecting)
state.
Initially the atoms reflect from the barrier as expected, but then the
atoms gradually pass through the barrier and become trapped on the
right-hand side.
This phenomenon is a direct result of choosing the main-barrier-beam
frequency to be more nearly resonant with the $F{=}2 \to F'$ transition
than the $F{=}1 \to F'$ transition.
Though the main barrier beam reflects many atoms during the initial
encounter, this interaction will also change the state of many of these
atoms from the $F{=}2$ reflecting state to the $F{=}1$ transmitting state.
When these atoms encounter the main barrier beam the second time, they will
transmit through, while atoms that did not change state on the first
reflection will eventually change their state upon subsequent reflections.
This results in a net accumulation of atoms on the right-hand side of the
barrier despite the atoms being initially in the ``wrong'' ($F{=}2$
reflecting) state.
The one-way barrier functions almost equally well for atoms on either side
of the barrier and for atoms in either state.

As a further test of the utility of the one-way barrier, we activated the
barrier after a ``filled-trap'' load of the dipole trap, which resulted in
a symmetric and relatively uniform initial distribution of atoms, all in
the $F{=}1$ (transmitting) ground state.
This experiment also had a lower barrier-beam power of $18(2) \ \mathrm{\mu
W}$, chosen to reduce heating without much harm to the function of the
barrier.
In this scheme atoms on the left-hand side of the barrier gradually
transmit to the right-hand side, while atoms on the right-hand side remain
there.
The results for this test are shown in Fig.~\ref{fig:waterfall}(e), where
many of the atoms that start to the left of the barrier eventually
transmit to the right-hand side.
Figure \ref{fig:compression} is generated from the same data, and shows the
populations on each side of the barrier as a function of time.
Here it is easier to see that atoms really are being transferred from the
left-hand side of the barrier to the right-hand side, in the manner of
Maxwell's demon.

\subsection{\label{subsec:scattering}Scattering}

\begin{figure}[tbp]
	\begin{center}
		\includegraphics[width=0.75\columnwidth]{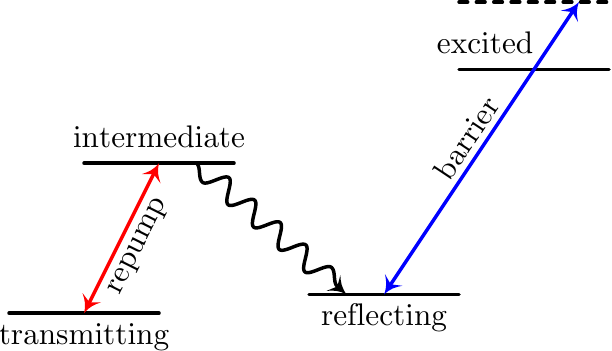}
	\end{center}
	\vspace{-5mm}
	\caption{%
		(Color online)
		Schematic of a four-level atom that would allow much larger
		barrier-beam detunings for reduced scattering.
		Here, the main barrier beam is detuned from a different transition than
		the one with which the repumping barrier beam is resonant.
		If the main-barrier-beam transition has a substantially different
		frequency than the repumping-barrier-beam transition, then the main
		barrier beam may be detuned by an amount greater than the ground-state
		splitting, and still create an effective barrier for atoms in the
		reflecting ground state while having negligible effect on atoms in the
		transmitting state.
	\label{fig:four-level}}
\end{figure}

The main technical challenge associated with implementing the one-way
barrier lies in the limited detunings afforded by the $6.8 \ \mathrm{GHz}$
ground-state hyperfine splitting of $^{87}\mathrm{Rb}$.
Detuning the main barrier beam halfway in between the $F{=}1$ and $F{=}2$
ground states seems an obvious choice; however, this is still near enough
to the $F{=}1 \to F'$ resonance to optically pump a portion of the atoms
(initially in the $F{=}1$ transmitting state) to the $F{=}2$ reflecting
state while they are traversing the barrier beam.
This results in a sudden increase in potential energy, greatly increasing
the total energy of the atom.
The heating associated with this transition is on the order of the barrier
height, which greatly decreases the effectiveness of the barrier.
This heating mechanism informed our choice of main-barrier-beam detuning,
in which the main barrier beam is more nearly resonant with the $F{=}2 \to
F'$ transition, with a $1.05(5) \ \mathrm{GHz}$ detuning from the $F{=}2
\to F'{=}3$ MOT trapping transition.
With this detuning, atoms in the $F{=}1$ ground state scatter few photons,
so heating during transmission is minimized.
Scattering during reflection is more probable because the main barrier beam
is closer to the resonances available to atoms in the $F{=}2$ ground state.
However, when atoms change state during reflection, their potential energy is
\emph{decreased}, which cools them.
Those atoms are then free to transmit, and are still cool enough for the
barrier to trap them when they return to the reflecting side.
Data collected to probe the effects of different main-barrier-beam
detunings are presented and analyzed in Sec.~\ref{sec:robustness}, where
we see that a more symmetric detuning does not function as well as this
asymmetric detuning.

For our canonical setup, which uses the detuning described above, we expect
$\sim\! 0.7$ scattering events on a single transmission and $\sim\! 8$
scattering events on a single reflection, ignoring state changes and any
scattering from the repumping barrier beam (see Appendix
\ref{app:analytic}).
We also ran simulations (see Sec.~\ref{sec:simulation} for more details)
that account for state changes, heating, and scattering from the repumping
barrier beam.
These simulations found slightly higher values of $\sim\! 3$ scattering
events for transmission and $\sim\! 10$ scattering events for a single
reflection.
In particular, note that even though the main barrier beam is detuned by
more than half of the hyperfine ground-state splitting for transmitting
atoms, scattering is not negligible.

Losses in general in this setup seem to be mainly due to light-assisted
collisions (discussed in Sec.~\ref{sec:simulation}) facilitated by the
barrier beams during reflection and transmission.
Light-assisted collisions include fine-structure-changing collisions
\cite{vigue1991}, hyperfine-changing collisions \cite{gould1995},
photoassociation \cite{heinzen1993}, and radiative escape
\cite{pritchard1989,wieman2000}.
We believe that, for our barrier beam, the dominant loss mechanism is
radiative escape \cite{pritchard1989,wieman2000}, although all can be
modeled as density-dependent loss mechanisms.
In radiative escape, an atom is promoted to an excited state by one of the
barrier beams.
While excited, the atom is much more polarizable and interacts more
strongly with neighboring atoms.
Before decaying, the atom accelerates toward a neighboring atom.
Once the atom has decayed back to the ground state, the atoms cease
interacting, but keep their kinetic energy.
This kinetic energy is often enough to eject both atoms from the trap.

Initial trap lifetimes range from $300$ to $500 \ \mathrm{ms}$ on the
right-hand (reflecting) side of the barrier depending on the temperature of
the atoms.
The fitted exponential lifetimes are markedly larger (approaching $700$ to
$900 \ \mathrm{ms}$) in the late-time tails of the population-decay curves.

The scattering rates are not too problematic in our setup, as the lifetimes
indicate.
However, the lifetimes could be improved considerably if we could reduce
the scattering rate by one or two orders of magnitude, so that the average
number of scattering events on transmission and reflection was well below
unity.
As shown in Appendix \ref{app:analytic}, the scattering rate depends mostly
on the barrier height, atomic speed, and the ratio of the linewidth to the
detuning.
Once the barrier is high enough to easily trap atoms, scattering during
reflection is only weakly dependent on barrier height.
Scaling down the atomic velocities limits the utility of the barrier as a
cooling mechanism, so increasing the detuning is perhaps the best variable
to use in reducing scattering.
A large increase in detuning could be achieved if the main barrier beam
used a transition with a substantially different frequency than any
available to the reflecting state (see Fig.~\ref{fig:four-level}).
This would allow the main barrier beam to be far-detuned from the
reflecting-to-excited state transition, and still several orders of
magnitude further detuned from the transmitting-to-intermediate state
transition.
Atoms in the transmitting state would thus be largely unaffected by the
main barrier beam because of the very large detuning, but could still be
pumped to the reflecting ground state by the repumping beam.
If necessary, a third laser beam detuned from the
transmitting-to-intermediate state could be used to cancel any shifts of
the transmitting state by the main barrier beam.
This would circumvent the ground-state-splitting limitation and lead to
perhaps orders of magnitude less scattering.

As a more specific example, we could use $^{88}\mathrm{Sr}$, with the true
ground state ($5s^2\phantom{a}^1\mathrm{S}_0$) as the transmitting state
and the metastable state ($5p\phantom{a}^3\mathrm{P}_2$) as the reflecting
state.
For the intermediate and excited states shown in Fig.~\ref{fig:four-level},
we could use $5p\phantom{a}^1\mathrm{P}_1$ and
$5d\phantom{a}^3\mathrm{D}_3$, respectively.
With these states, the main-barrier-beam transition is almost $40\
\mathrm{nm}$ red of the repumping-barrier-beam frequency and has a
linewidth comparable to the $^{87}\mathrm{Rb}$ $\mathrm{D}_2$ transitions.
This would allow the main barrier beam to be detuned much further than the
$^{87}\mathrm{Rb}$ ground-state splitting and still be many nanometers
detuned from the repumping-barrier-beam transition.
This could reduce scattering during reflection by multiple orders of
magnitude and reduce scattering during transmission to negligible levels.
The repumping decays presented here are slow (millisecond time scales) and
involve other states that may decay to either the reflecting state or back
to the transmitting state, but all routes back to the transmitting state
are short compared with the reflecting state lifetime.
The long repumping time could be corrected for by allowing the repumping
barrier beam to fill the entire trapping side of the barrier, allowing for
more interaction time.

\section{\label{sec:robustness}Robustness}

To investigate the robustness of the one-way barrier, we varied four key
parameters in the experiment:
\begin{enumerate}
	\item
	the separation between the main barrier beam and the repumping barrier
	beam, from $8$ to $74 \ \mathrm{\mu m}$;
	\item
	the initial displacement of the atoms (this represents an effective
	change in barrier height), from $0$ to $0.9 \ \mathrm{mm}$ on either side
	of the barrier;
	\item
	the power of the repumping barrier beam, from $0.002$ to $8.7 \
	\mathrm{\mu W}$; and
	\item
	the detuning of the main barrier beam from the $F{=}2$ ground state of
	$^{87}\mathrm{Rb}$, from $+0.75$ to $+4 \ \mathrm{GHz}$ (blue-detuned),
	measured with respect to the $^{87}\mathrm{Rb}$ MOT trapping transition.
\end{enumerate}

\begin{figure*}[tbp]
	\begin{center}
		\includegraphics[width=\textwidth]{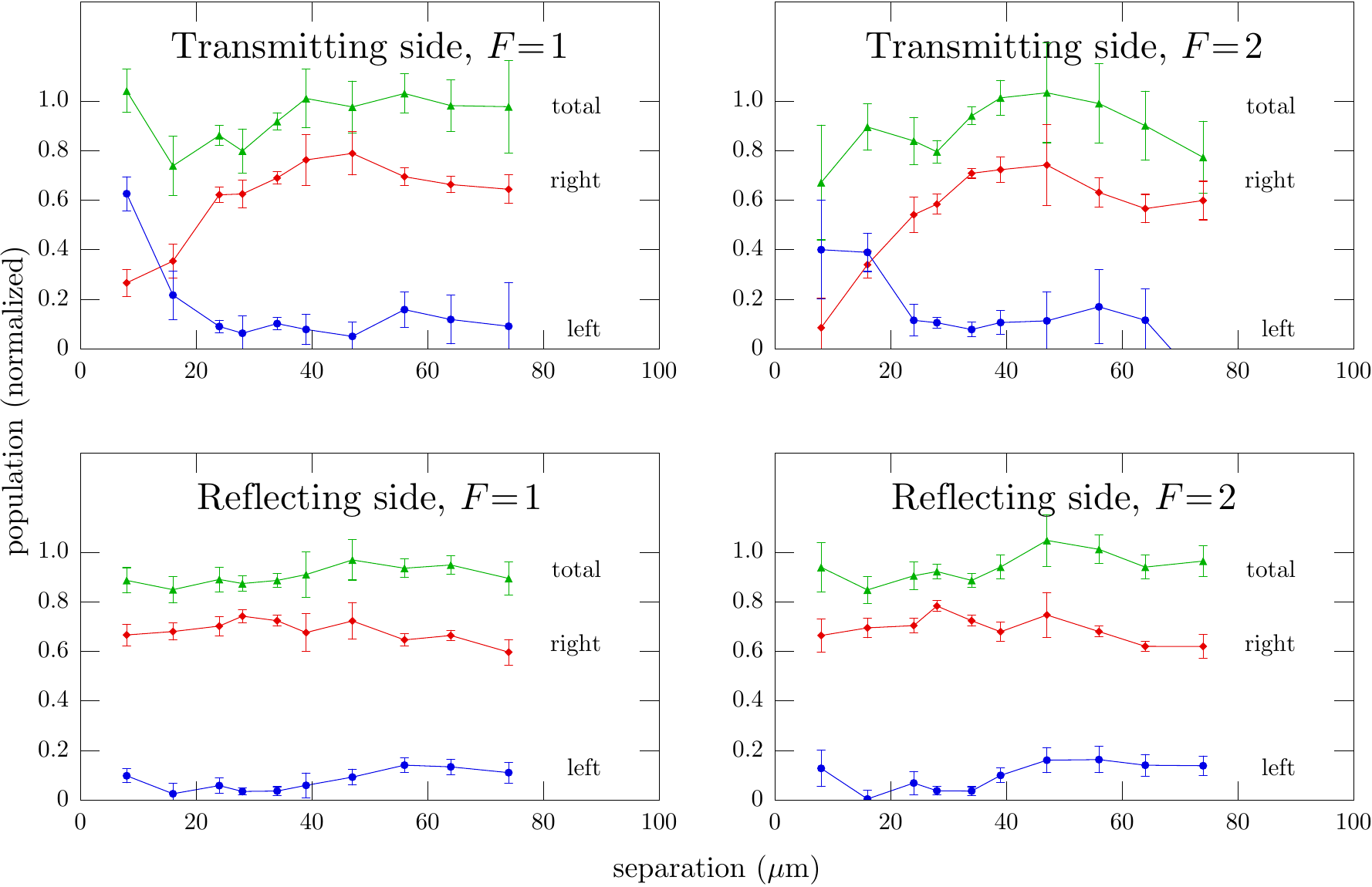}
	\end{center}
	\vspace{-5mm}
	\caption{%
		(Color online)
		Populations on the left- and right-hand sides of the one-way barrier
		after $100 \ \mathrm{ms}$ as functions of barrier-beam separation.
		From left to right, top to bottom, atoms were initially loaded on the
		left in the $F{=}1$ ground state, on the left in the $F{=}2$ ground
		state, on the right in the $F{=}1$ ground state, and on the right in
		the $F{=}2$ ground state.
		Error bars indicate statistical error from at least $20$
		repetitions.
	\label{fig:beam_sep}}
\end{figure*}

Initially the separation between the beams was set to
$34 \ \mathrm{\mu m}$, which ensured that the tails of the Gaussian beam
profiles along the dipole-trap axis overlapped slightly.
We believed the overlap was necessary to prevent the atoms trapped on the
right-hand side of the barrier in the $F{=}2$ reflecting state from being
pumped back to the $F{=}1$ transmitting state by the main barrier beam
(with an effective two-level $\Delta \approx 2\pi \times 1.12 \
\mathrm{GHz}$), such that they would pass back through the barrier to the
transmitting side.
The presence of a small amount of repumping-barrier-beam light in the same
region would rapidly pump any atoms that made it to the $F{=}1$
transmitting state back to the reflecting $F{=}2$ state.
The overlap was designed to guarantee that atoms did not slowly leak back
across the barrier.

Surprisingly, our studies of the effect of barrier-beam separation indicate
that while overlapping the beams is important for optimizing barrier
performance, it is not critical to barrier operation.
Figure \ref{fig:beam_sep} shows populations on either side of the barrier
after $100 \ \mathrm{ms}$ of evolution as functions of beam separation.
The transmitting-side data show the most dramatic effect for separations
below about $30 \ \mathrm{\mu m}$.
When the beams are too close together, the tail of the repumping
barrier beam can pump atoms on \emph{either} side of the barrier to the
reflecting state, causing the barrier to reflect atoms from both sides.
This can be seen for the small separations (on the order of two beam waists
and below) on the transmitting-side data, where the majority of atoms
stayed on the left-hand (supposedly transmitting) side of the barrier.
We also see a higher loss rate for small separations, which we suspect may
be due to atoms changing state in the middle of the main barrier beam as
opposed to the tail, as well as changing state more frequently.
When many state changes occur in the middle of the barrier beam, the large
changes in potential energy result in a large amount of heating, which
causes much more loss.
Large separations show much less effect on the transmitting-side data,
although the $F{=}2$ data seem to get worse and then improve as the
separation increases past $50 \ \mathrm{\mu m}$.
For atoms beginning to the right of the barrier, the beam separation does
not severely affect the barrier's reflectivity, though the functionality of
the barrier declines as the separation is increased beyond about $50 \
\mathrm{\mu m}$.
As long as the repumping barrier beam is still present to the right of the
main barrier beam, most atoms will pass back through the repumping
barrier beam after reflection, correcting any state changes that may have
occurred.
As the separation between the beams increases, more atoms may fail to make
it back to the repumping barrier beam after multiple reflections,
explaining the slow reduction in reflectivity beyond $50 \ \mathrm{\mu m}$.
This observation has implications for one-way barrier cooling
applications that require a significant number of reflections from
the barrier.

We investigated how the loading position of the atoms in the dipole trap
affected the dynamics of the one-way barrier as a substitute for changing
the barrier height.
The initial position affects the ratio of average atomic kinetic energy to
barrier height, and so makes a decent substitute for varying the barrier
height.
However, it is important to note that changing the kinetic energy
without adjusting the barrier height is not quite the same as
changing the barrier height without altering the kinetic energy;
the two changes have different effects on the expected number of
scattering events while the atoms traverse the barrier.
Appendix \ref{app:analytic} contains a more detailed analysis of this
assertion.

\begin{figure}[tbp]
  \begin{center}
	  \includegraphics[width=\columnwidth]{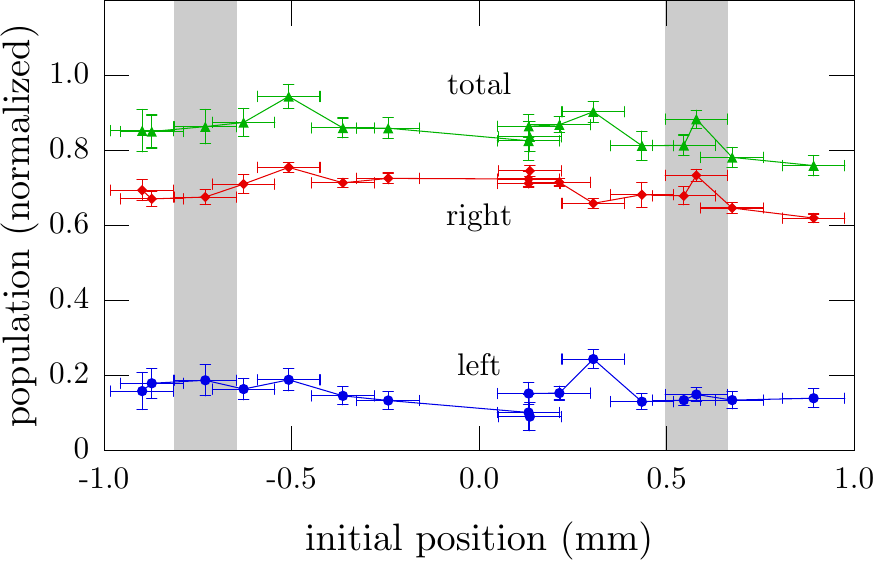}
	\end{center}
	\vspace{-5mm}
	\caption{%
		(Color online)
		Populations on the left- and right-hand sides of the one-way barrier
		after $100 \ \mathrm{ms}$ as functions of initial loading position.
		The initial loading position is the position of the zero of the MOT
		magnetic fields relative to the focus of the dipole trap, in
		millimeters.
		Our canonical starting locations were about $0.65(8) \ \mathrm{mm}$
		away from the focus of the dipole trap and are marked by the two
		vertical gray regions.
		Vertical error bars indicate statistical error from $38$ repetitions,
		and horizontal error bars indicate uncertainty in measuring the MOT
		center.
	\label{fig:pull_back}}
\end{figure}

\begin{figure}[tbp]
	\begin{center}
		\includegraphics[width=\columnwidth]{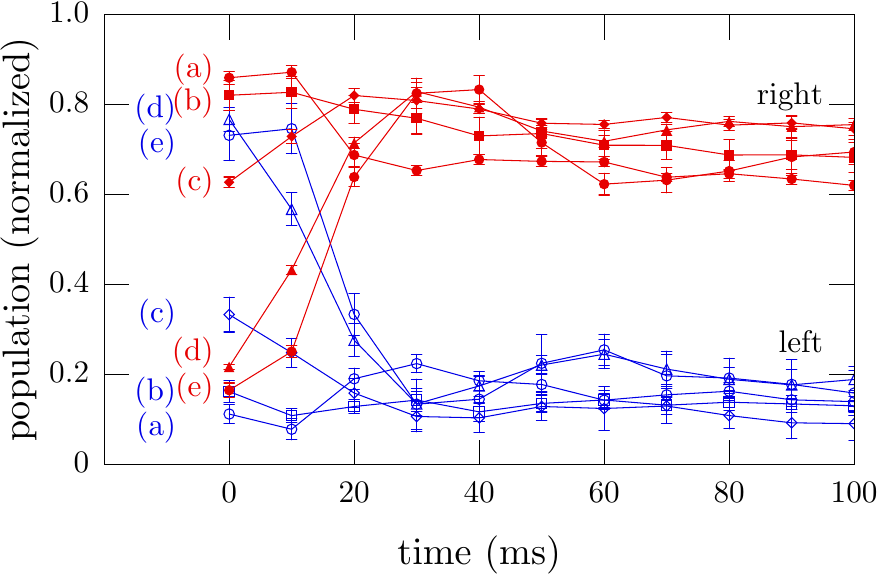}
	\end{center}
	\vspace{-5mm}
	\caption{%
		(Color online)
		Some sample time-series that formed the data in
		Fig.~\ref{fig:pull_back}, showing populations on the right- and
		left-hand sides of the barrier as functions of time.
		Shown are curves for initial loading positions of (a) $0.89(8) \
		\mathrm{mm}$, (b) $0.43(8) \ \mathrm{mm}$, (c) $0.14(8) \ \mathrm{mm}$,
		(d) $-0.51(8) \ \mathrm{mm}$, and (e) $-0.90(8) \ \mathrm{mm}$ from the
		barrier center.
		Error bars indicate statistical error from $38$ repetitions.
	\label{fig:pull_back_time}}
\end{figure}

In Fig.~\ref{fig:pull_back}, we plot the atomic populations in the one-way
barrier at $100 \ \mathrm{ms}$ as functions of initial loading position, as
determined by the location of the center of the MOT trapping fields.
We see no obvious effect of the loading position on the efficiency of the
one-way barrier.
Figure \ref{fig:pull_back_time} shows some sample time-series for various
initial loading positions.
Essentially, we see that atoms starting on the transmitting side are
transferred to the reflecting side, and atoms starting on the reflecting
side are kept there, independently of initial loading position.
This indicates that the efficiency of our one-way barrier is fairly
insensitive to the velocity of the incident atoms.

\begin{figure*}[tbp]
	\begin{center}
		\includegraphics[width=\textwidth]{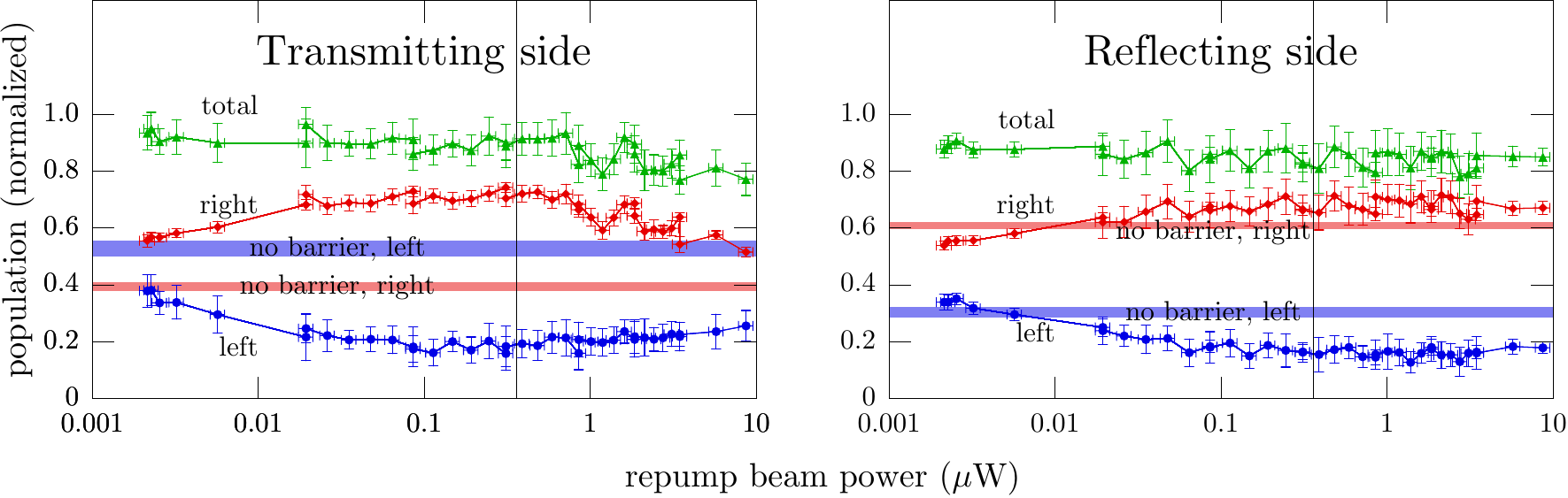}
	\end{center}
	\vspace{-5mm}
	\caption{%
		(Color online)
		Populations on the left- and right-hand sides of the one-way barrier
		after $100 \ \mathrm{ms}$ as functions of repumping-barrier-beam power.
		The left panel shows results where the atoms were initially on the
		left-hand (transmitting) side of the barrier, and the right panel shows
		results where the atoms were initially on the right-hand (reflecting)
		side.
		The vertical lines in the plots show our usual repumping-barrier-beam
		power of $0.36 \ \mathrm{\mu W}$.
		The two horizontal bars show the ending populations from data taken
		with no barriers [the one shown in column (a) of
		Fig.~\ref{fig:waterfall} and a similar data set where atoms were
		initially on the other side of the barrier].
		The no-barrier cases show unequal left- and right-hand populations
		due to the oscillations of the atoms.
		Error bars indicate statistical error from at least $38$
		repetitions.
	\label{fig:repump}}
\end{figure*}

\begin{figure*}[tbp]
	\begin{center}
		\includegraphics[width=\textwidth]{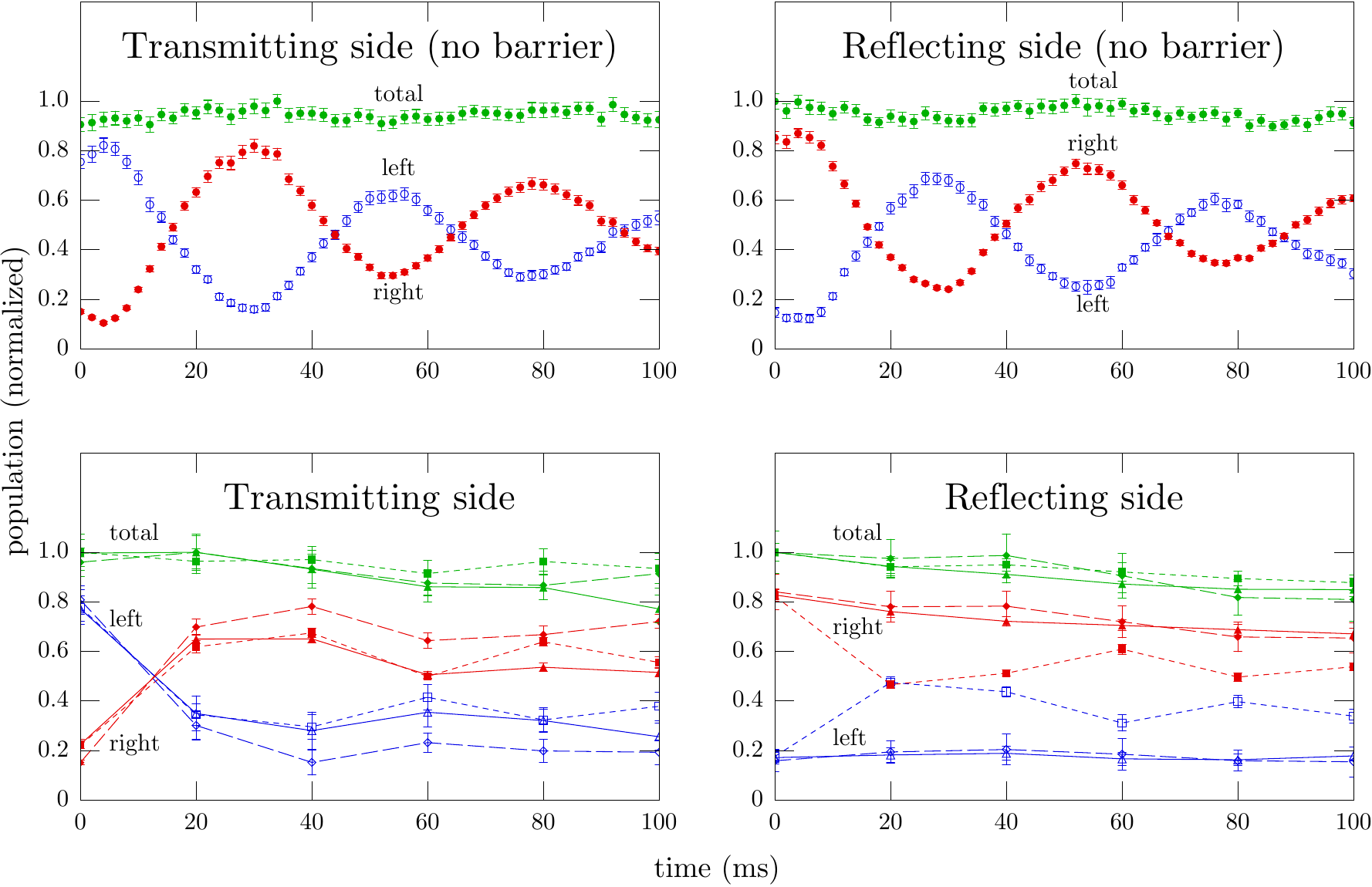}
	\end{center}
	\vspace{-5mm}
	\caption{%
		(Color online)
		Populations on the left- and right-hand sides of the one-way barrier
		for a few different repumping-barrier-beam powers as functions of time.
		The top two panels show data collected without barrier beams.
		The bottom two panels are generated from the same data used in
		Fig.~\ref{fig:repump}.
		The data with the dotted line and squares correspond to the minimum
		beam power (about $2 \ \mathrm{nW}$) shown in Fig.~\ref{fig:repump}.
		The data with the dashed line and diamonds correspond to approximately
		the same power used to collect our canonical data.
		The data with the solid line and triangles correspond to the maximum
		beam power (about $9 \ \mathrm{\mu W}$) shown in Fig.~\ref{fig:repump}.
		The top curve in each panel shows the total population, while the other
		solid symbols show the right-hand side population and the open symbols
		show the left-hand side population.
	\label{fig:repump_time}}
\end{figure*}

While the intensity of the main barrier beam controls the height of the
barrier, the intensity of the repumping barrier beam also plays a crucial
role in the transmission and reflection of atoms from the barrier.
We varied the intensity of the repumping barrier beam by over three orders
of magnitude to observe its effect on the functioning of the barrier.
The results for atoms that started on the left-hand and right-hand sides of
the barrier are presented in Fig.~\ref{fig:repump}, where the populations
of atoms on either side of the barrier after $100 \ \mathrm{ms}$ are
measured for different repumping-barrier-beam intensities.
Even over the full range of intensities we studied, the barrier continued
to act asymmetrically in both experiment and simulations.
This can be seen in Fig.~\ref{fig:repump_time}, which has time-series plots
of the highest and lowest repumping-beam intensities we studied.
The barrier continues to operate asymmetrically, although the efficiency is
strongly affected.

The transmitting-side data in Fig.~\ref{fig:repump} show the expected weak
optimum in repumping-barrier-beam intensity.
When the repumping barrier beam is too weak to pump atoms efficiently to
the reflecting state, the atoms continue to transmit through the barrier,
resulting in very little asymmetry, save for whatever side the atoms happen
to be on after $100 \ \mathrm{ms}$.
However, even at the weakest intensity used in Fig.~\ref{fig:repump}, the
repumping barrier beam still pumped atoms to the reflecting state and the
barrier acted asymmetrically, although less efficiently.
As the intensity of the repumping barrier beam is increased, the tails of
its beam profile, which extend to the left (transmitting side) of the main
barrier beam, will eventually have enough intensity to optically pump atoms
on the left-hand side of the barrier into the reflecting state before they
reach the main barrier beam.
In this limit the barrier becomes reflective on both sides, such that all
atoms will remain to the left of the barrier.
Though we do not reach this limit in our experiment, these two limits
together indicate the existence of an optimum repumping-barrier-beam
intensity that we do observe in Fig.~\ref{fig:repump}.

For the reflecting-side data in Fig.~\ref{fig:repump}, we see the same
decrease in effectiveness as the repumping barrier beam becomes too weak to
pump atoms to the reflecting state.
However, we do not see the same decrease in barrier efficiency as the
repumping-barrier-beam intensity is increased, since a stronger repumping
barrier beam just causes the barrier to reflect better.
These behaviors in the high- and low-intensity limits are evident in the
reflecting-side data shown in Fig.~\ref{fig:repump}, where the barrier
ceases to effectively block atoms as the intensity decreases but its
performance improves and then saturates (presumably at the point where all
atoms reflect) as the intensity increases.
At $t=0$, the atoms in the dipole trap have just enough spread that the
tails of the distribution cross the barrier, as can be seen at $t=0$ in
Fig.~\ref{fig:repump_time}.
This is a likely explanation as to why the population on the left-hand side
of the barrier is not zero in Fig.~\ref{fig:repump}, even for the largest
repumping-barrier-beam intensities.

\begin{figure}[tbp]
	\begin{center}
		\includegraphics[width=\columnwidth]{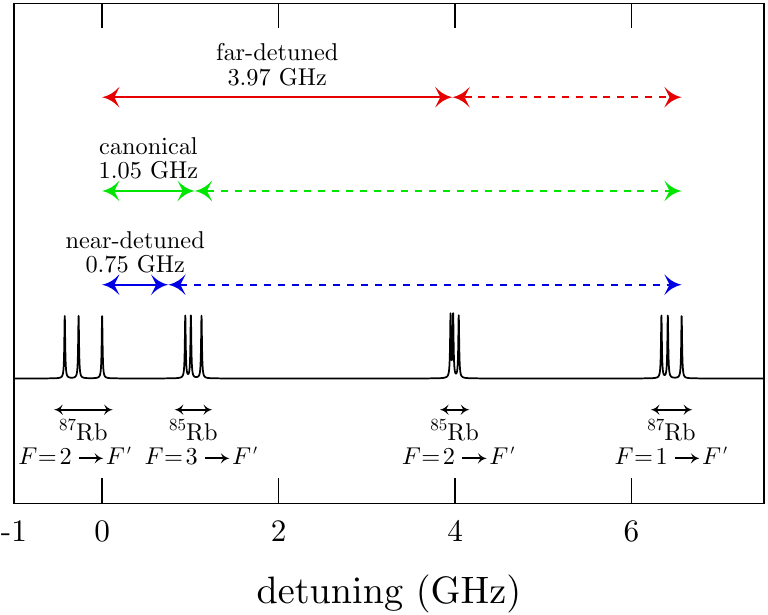}
	\end{center}
	\vspace{-5mm}
	\caption{%
		(Color online)
		A schematic of the rubidium emission spectrum around the $\mathrm{D}_2$
		line, showing the detunings we tested for the main barrier laser.
		The horizontal axis measures detuning from the $^{87}\mathrm{Rb}$
		$F{=}2 \to F'{=}3$ MOT trapping transition.
		Both $^{87}\mathrm{Rb}$ and $^{85}\mathrm{Rb}$ resonances are shown.
		Our canonical detuning coincides with the $^{85}\mathrm{Rb}$ $F{=}3 \to
		F'{=}3,4$ crossover transition, and the far-detuned value coincides
		with the $^{85}\mathrm{Rb}$ $F{=}2 \to F'{=}2$ transition.
		For the near-detuned value, we monitored the detuning with a
		Fabry-P\'erot cavity.
		Each detuning from the $^{87}\mathrm{Rb}$ MOT trapping transition is
		shown with a solid arrow.
		This approximates the detuning seen by atoms in the $F{=}2$
		(reflecting) state.
		A dotted arrow shows the detuning from the $^{87}\mathrm{Rb}$ MOT
		repumping transition, which approximates the detuning seen by atoms in
		the $F{=}1$ (transmitting) state.
	\label{fig:rb_spectrum}}
\end{figure}

\begin{figure}[tbp]
	\begin{center}
		\includegraphics[width=\columnwidth]{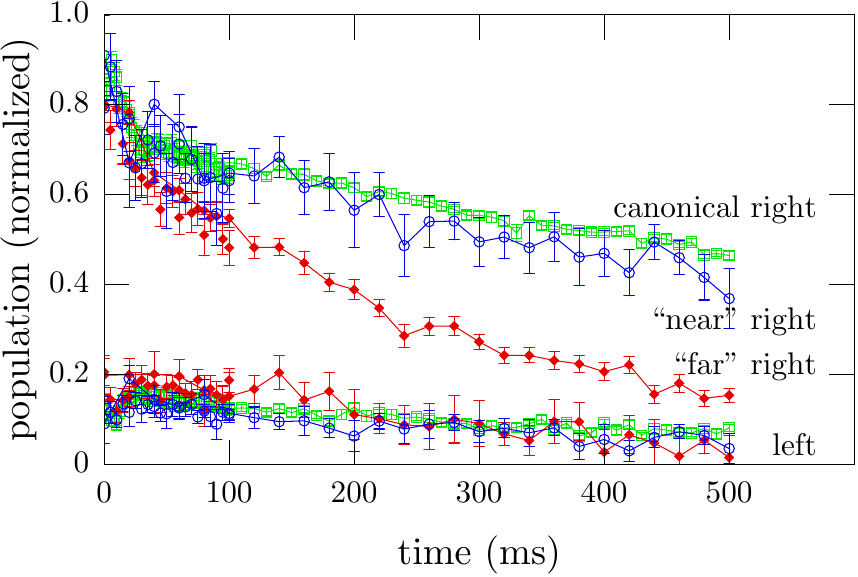}
	\end{center}
	\vspace{-5mm}
	\caption{%
		(Color online)
		Populations on the left- and right-hand sides of the one-way barrier,
		comparing barrier performances when the main barrier laser operates at
		different frequencies.
		For the canonical, near-detuned, and far-detuned data, the main-barrier
		detunings were $1.05(5) \ \mathrm{GHz}$, $0.75(5) \ \mathrm{GHz}$, and
		$3.97(7) \ \mathrm{GHz}$ blue of the $^{87}\mathrm{Rb}$ MOT transition,
		respectively.
		See Fig.~\ref{fig:rb_spectrum} for details.
		The near-detuned and canonical data are very similar, while the
		far-detuned data show much more atom loss.
		Error bars indicate statistical error from at least $28$
		repetitions.
	\label{fig:freq_life}}
\end{figure}

We measured population losses for atoms released on the right-hand side of
the barrier for two different main-barrier-beam detunings (shown in
Fig.~\ref{fig:rb_spectrum}).
Figure \ref{fig:freq_life} shows the results.
The near-detuned and far-detuned data were collected with the main barrier
beam tuned $0.75(5) \ \mathrm{GHz}$ (using a Fabry-P\'erot cavity) and
$3.97(7) \ \mathrm{GHz}$ (using the $^{85}\mathrm{Rb}$ $F{=}2 \to F'{=}2$
transition) blue of the $^{87}\mathrm{Rb}$ MOT trapping transition,
respectively.
The near-detuned value was picked to be substantially closer to the $F{=}2
\to F'$ resonances than our canonical value, but not so close as to
increase scattering of the barrier-beam light enough to prevent barrier
operation.
The far-detuned value was conveniently close to halfway between the
resonances for $F{=}1$ (transmitting) and $F{=}2$ (reflecting) atoms, and
is actually a little closer to the $F{=}1$ resonances.
This allowed us to compare a more symmetric detuning to our canonical case
and test how damaging state-changing scattering events are to the barrier
operation.
For both of these data sets, the intensity of the main barrier beam was
adjusted so that the height of the reflecting barrier was the same as for
our canonical data (with the barrier tuned to the $^{85}\mathrm{Rb}$ MOT
transition), which is also shown in Fig.~\ref{fig:freq_life}.
The near-detuned data are nearly indistinguishable from our canonical
reflecting-side data.
The far-detuned data show a much shorter lifetime, but no extra leakage to
the reflecting side of the barrier.
This indicates the following:
\begin{enumerate}
	\item
	Heating from scattering barrier-beam light alone is not much of a
	problem.
	If it were, then the near-detuned data, which should have $\sim\! 2$
	times as much scattering as our canonical data because it is closer to
	resonance, should show much more loss.
	\item
	State-changing from scattering barrier-beam light \emph{is} an issue.
	We believe we can explain this as follows.
	In both our canonical data and the near-detuned data, the barrier was
	much more likely to pump atoms to the transmitting state, which
	\emph{decreases} the potential energy of the atoms, thus helping to cool
	them.
	The repumping barrier beam pumps to the reflecting state, but since that
	occurs away from the barrier beam, heating due to the potential-energy
	increase is small.
	In the far-detuned data, the main barrier beam can pump atoms to either
	state.
	Atoms in the transmitting state that are pumped to the reflecting state
	are likely to be ejected from the barrier at high speed.
	Atoms in the reflecting state that are pumped to the transmitting state
	are more likely to be moving slowly, because they were being reflected,
	which makes them likely to scatter again.
	Thus, atoms leaving the beam after a state change are more likely to have
	been heated than cooled.
	This increased heating decreases the trap lifetime substantially.
	Appendix \ref{app:analytic} discusses scattering in more depth.
\end{enumerate}

We also briefly tried having both the main barrier beam and the repumping
barrier beam polarized parallel to the longitudinal axis of the dipole-trap
beam.
We saw no significant change, suggesting that the barrier operation is
insensitive to polarization.

\section{\label{sec:simulation}Simulations}

We simulated the barrier operation using a simple model where atoms were
assumed to be noninteracting point particles with well-defined positions
and momenta.
These atoms move in conservative potentials formed by the dipole-trap beam
and the main barrier beam.
At every point in time the atoms were assumed to be in either the $F{=}1$
(transmitting) or $F{=}2$ (reflecting) ground state.
This is because coherences between these states oscillate at much faster
time scales (subnanosecond) than the center-of-mass motion (microseconds
to milliseconds) of the atoms, and so can be time averaged, effectively
allowing each atom to be in a classical mixture of the two ground states.

The potential the atoms move in is a sum of potentials from the dipole-trap
beam and the main barrier beam.
We assume the strengths of the potentials to be proportional to the local
beam intensities.
We use far-detuned approximations for computing the potentials, and we also
use the rotating-wave and two-level-atom approximations for the barrier
beams.
The repumping barrier beam pumps atoms out of the $F{=}1$ ground state very
quickly, and is too weak and too far detuned to produce a strong potential
for atoms in the $F{=}2$ ground state, so we ignore its potential.
After each time step of the integration, we compute a local scattering rate
for each atom given the beam intensities at the atom's position and the
current atomic state.
The dipole-trap beam is very far detuned and has a scattering rate on the
order of $3\ \mathrm{s}^{-1}$, which we ignore---in our simulation only the
barrier beams scatter off the atoms.
We then randomly decide whether each atom scatters a photon in this time
step, with a probability equal to the scattering rate times the time step.
A scattering event is modeled by a randomly directed (with a
dipole-emission distribution) single-photon recoil kick applied to the
momentum, and a randomly chosen atomic state change based on the
probabilities computed in Appendix \ref{app:state_change}.

\begin{figure*}[tb]
	\begin{center}
		\includegraphics[width=\textwidth]{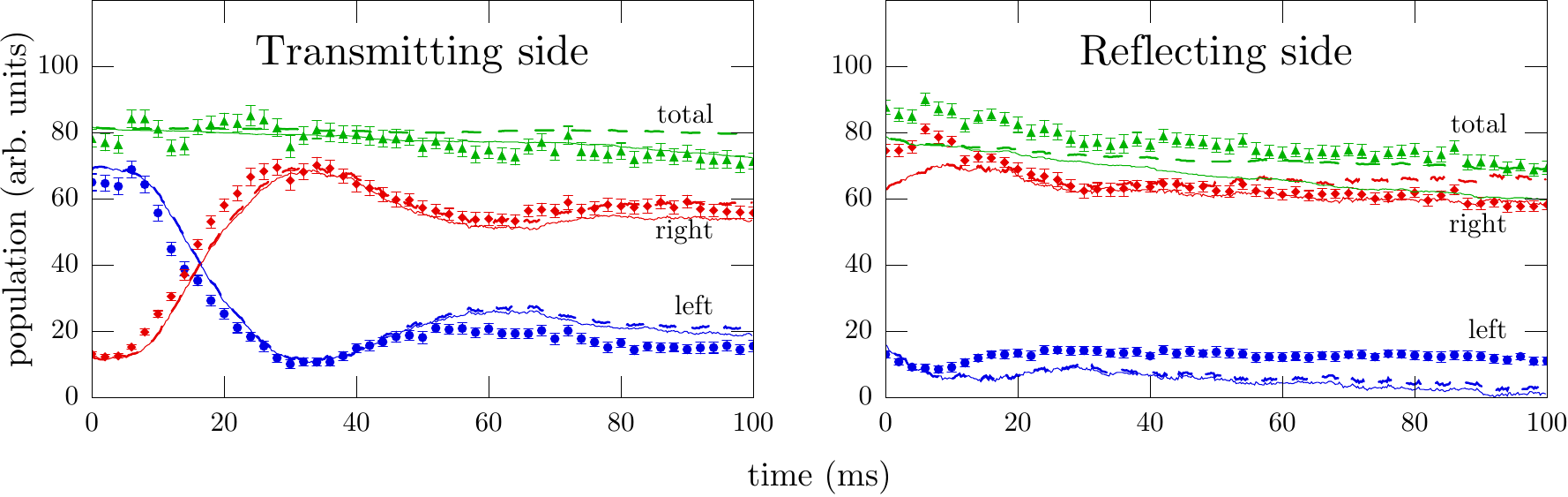}
	\end{center}
	\vspace{-5mm}
	\caption{%
		(Color online)
		Populations on the left- and right-hand sides of the one-way barrier,
		comparing data (points) to simulations (solid and dashed lines).
		The solid simulation curve has an extra light-assisted collisional
		loss mechanism.
	\label{fig:sim_left_right}}
\end{figure*}

\begin{figure*}[tb]
	\begin{center}
		\includegraphics[width=\textwidth]{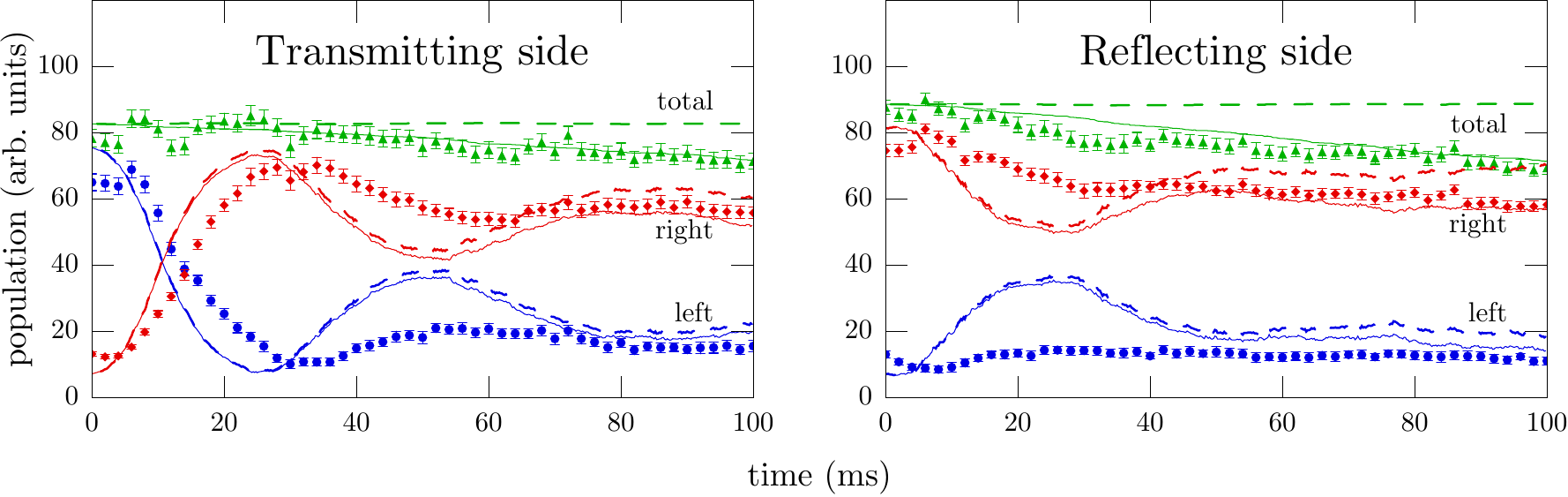}
	\end{center}
	\vspace{-5mm}
	\caption{%
		(Color online)
		Populations on the left- and right-hand sides of the one-way barrier,
		comparing data (points) to simulations (solid and dashed lines).
		These simulations and data are the same as in
		Fig.~\ref{fig:sim_left_right}, except that the simulation initial
		conditions had no transverse spread in either position or momentum.
		These simulations do not match the data as well as the simulations in
		Fig.~\ref{fig:sim_left_right}, illustrating how basic barrier results,
		such as the time atoms take to reach the barrier and whether atoms
		oscillate across the barrier, are strongly affected by the initial
		conditions.
		The solid simulation curve has an extra light-assisted collisional
		loss mechanism.
	\label{fig:sim_left_right_1D}}
\end{figure*}

We found that the simulations were somewhat sensitive to initial conditions
(compare Fig.~\ref{fig:sim_left_right} with
Fig.~\ref{fig:sim_left_right_1D}).
We achieved very good agreement between simulation and experiment with the
following initial conditions:
Atoms were initially placed in an oblong Gaussian ellipsoid, placed to
match the initial conditions of the atoms in the experiment we were
modeling.
The long axis lies along the dipole-trap axis, with a length comparable to
that measured in our experiments (we used a standard deviation of $400 \
\mathrm{\mu m}$).
The two short-axis widths were chosen so that most of the atoms started out
trapped in the dipole trap, with some small variations to help match the
simulations to the data.
The momenta for each atom were initialized to random values with a
$\sim\! 100 \ \mathrm{\mu K}$ Maxwell-Boltzmann distribution.
We also simulated the nonzero loading time.
For each atom we picked a uniformly distributed random start time that was
less than the loading time.
Each atom is frozen until its chosen start time, at which point its
position and momentum begin to change according to the rules described
above.
This emulates the actual loading procedure, where atoms enter the
dipole-trap beam and become trapped throughout the loading time.
Atoms just entering the trapping region do so with random momenta because
atoms in a MOT scatter often.
Once trapped in the dipole trap they become dark to the MOT beams and
begin evolving according to the dipole-trap potential.
Thus, atoms that have spent time in the trap have a preferred direction and
position different from atoms that have just been loaded, as they have
begun moving toward the dipole-trap focus.
This gives a slight position-momentum correlation that just-loaded atoms do
not have.
The intention of having a load time in the simulation was to model this
correlation.

We also noticed that atoms loaded differently on one side of the barrier
than on the other.
This is likely due to fringes in the MOT trapping beams.
Thus, in our simulations, we used two slightly different sets of initial
conditions.
Atoms loaded on the transmitting side of the barrier had a temperature of
$100 \ \mathrm{\mu K}$, with a transverse standard deviation of $16 \
\mathrm{\mu m}$.
Atoms loaded on the reflecting side of the barrier had a temperature of $80
\ \mathrm{\mu K}$, with a transverse standard deviation of $8 \ \mathrm{\mu
m}$.
In both cases, the atomic positions were centered about a point on the trap
axis, $0.9 \ \mathrm{mm}$ from the dipole-trap focus and the barrier beams.

Figure \ref{fig:sim_left_right} was derived from the same data as columns
(b) and (d) of Fig.~\ref{fig:waterfall}, but with simulation results added
in for comparison.
The dashed simulation curve shows the results of the simulations as
described so far.
As an example of how the simulations are sensitive to initial conditions,
Fig.~\ref{fig:sim_left_right_1D} shows the same experimental data as
Fig.~\ref{fig:sim_left_right} compared to different simulations where the
initial conditions had the same longitudinal spreads but \emph{zero}
transverse position and momentum spreads.
These initial conditions have \emph{less} kinetic energy than those used in
Fig.~\ref{fig:sim_left_right}, yet atoms pass through the barrier much more
effectively.
This is caused by an orbital effect in the dipole-trap beam, where atoms
with angular momentum are slowed in their progression toward the trap focus
(see Appendix \ref{app:ang_mom}).
As can be seen by comparing Fig.~\ref{fig:sim_left_right} with
Fig.~\ref{fig:sim_left_right_1D}, the presence of angular momentum
(Fig.~\ref{fig:sim_left_right}) does effectively slow the atoms.
In some of our simulations, this slowing accounts for some of the atoms
remaining on the transmitting side of the barrier after $100 \
\mathrm{ms}$, because they approach the barrier too slowly to even reach it
on that time scale.
This effect can also be seen in our experimental data (i.e.,\
Fig.~\ref{fig:compression}).

\begin{figure}[tbp]
	\begin{center}
		\includegraphics[width=\columnwidth]{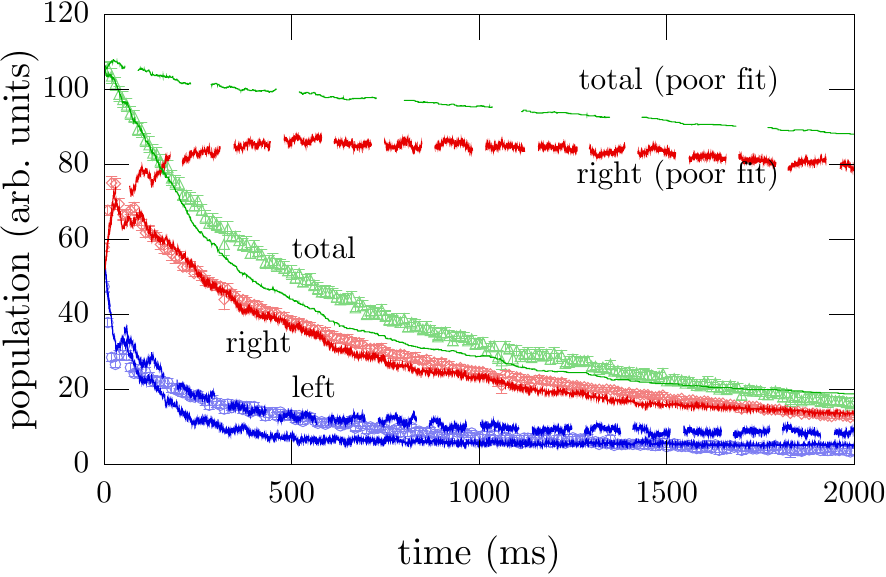}
	\end{center}
	\vspace{-5mm}
	\caption{%
		(Color online)
		Populations on the left- and right-hand sides of the one-way barrier,
		comparing data (points) to simulations (solid and dashed lines) for a
		long experiment showing the barrier lifetime.
		The trap was initially loaded nearly symmetrically.
		The dashed curve shows a simulation without including light-assisted
		collisions, which models the data very poorly.
		The solid curve shows the same simulations including light-assisted
		collisions, which matches the data very closely.
		For each curve (data and both simulations), the curves from top to
		bottom show the total population, the population on the right side
		of the barrier, and the population on the left side of the barrier.
		Error bars indicate statistical error from $19$ repetitions.
	\label{fig:sim_long}}
\end{figure}

Figure \ref{fig:sim_long} shows data taken from an experiment much like the
one in Fig.~\ref{fig:compression}, except with a slightly lower
barrier-beam power and carried out over much longer time scales.
The simulations as described (dashed line) completely miss the
experiment---we needed to include light-assisted collisional losses to best
fit the data.
In cold-atom traps there are several mechanisms where light affects
collisions between atoms
\cite{pritchard1989,vigue1991,heinzen1993,gould1995,wieman2000}, which we
collectively refer to as light-assisted collisions.
In all of these an incident laser beam excites an atom which then decays.
While the atom is excited, it either is affected by a different atom--atom
potential or decays to a different state through an interaction with
another atom, resulting in a large change in kinetic energy that ejects the
atom (and maybe another) from the trap.
Thus, all of these loss mechanisms may be collectively modeled as a
density-dependent loss term with a coefficient that is proportional to the
probability that atoms may be excited, although we believe that one in
particular (radiative escape \cite{wieman2000}) is responsible for our
losses.

We modeled light-assisted collisions by estimating the density of atoms in
the barrier region and randomly removing them with a loss rate proportional
to that density (and the beam intensity).
More precisely, if we let $N$ be the number of atoms in a small volume $V$,
then we expect the loss equation for $dN/dt$ to contain a term of the form
$cIN^2/V\Delta^2$, where $I$ is the local laser intensity, and $\Delta$ is
the detuning of the beam.
Then $I/\Delta^2$ is approximately proportional to the scattering rate of
the atoms and the $c$ coefficient is independent of volume.
This is the loss term we used to model light-assisted collisions.
We picked $c \sim 3 \times 10^{5} \ \mathrm{cm}^5 \, \mathrm{mW}^{-1} \,
\mathrm{s}^{-3}$ to match the data shown in Fig.~\ref{fig:sim_long}, and
used that value in all other simulations.

The difference between simulations with and without light-assisted
collisions can be seen in Fig.~\ref{fig:sim_long}, where the simulation
without light-assisted collisions is a very poor fit for the data, but the
simulation with light-assisted collisions can be made to fit the data very
well.
The differences can also be seen in Fig.~\ref{fig:sim_left_right}.
They are not as pronounced on the shorter times scales, but the simulations
with light-assisted collisions still model the data better.
This suggests that light-assisted collisions are the dominant loss
mechanism for our one-way barrier.
Density-independent effects provide a decent (exponential) fit to the data
in Fig.~\ref{fig:sim_long}, but we know of no mechanism that could explain
such a large loss rate.
We can rule out any effect not directly related to the barrier beams, such
as collisions with background atoms, due to the much longer lifetime of
atoms in the dipole trap (about $20 \ \mathrm{s}$) without the barrier
beams.

If we divide our $c$ coefficient by the appropriate detuning (squared) for
the experiment done by Kuppens \textit{et~al.}\ \cite{wieman2000}, we find
a coefficient of about $2 \times 10^{-9} \ \mathrm{cm}^5 \,
\mathrm{mW}^{-1} \, \mathrm{s}^{-1}$.
This should compare to the $K$ value of $1.1(5) \times 10^{-10} \
\mathrm{cm}^5 \, \mathrm{mW}^{-1} \, \mathrm{s}^{-1}$ quoted in
Ref.~\cite{wieman2000}.
Our density-dependent loss simulations were too simplistic and approximate
to expect better than order-of-magnitude agreement.

\section{\label{sec:maxwell}Connection to Maxwell's Demon}

In his 1871 \textit{Theory of Heat}, Maxwell \cite{maxwell1871}
contemplated a hypothetical creature who, being able to distinguish
individual molecules in a gas, could separate hot molecules from cold
molecules, resulting in a heat flow from a cold temperature to a hot
temperature.
This creature became known as Maxwell's demon, and its apparent ability to
violate the second law of thermodynamics remained an unsolved quandary for
decades.
Everyday experiences, and even the thermodynamical definition of
temperature, dictate that heat flows from high to low temperatures in an
\emph{irreversible} fashion.
This implies that separating hot and cold must \emph{decrease} the entropy
of a system.
One can also see how such a demon could be used to perform useful work by
creating a temperature differential and using an engine to extract energy
from that.

Many variations of Maxwell's demon have been proposed, including a simpler
demon that allowed particles into part of a larger chamber, but not back
out again---a one-way barrier \cite{bennett1987}.
Compressing the physical volume of a gas without changing the temperature
also decreases entropy, and again allows for work extraction from a heat
reservoir without the need for heat flow into a colder reservoir.

Today, Maxwell's demon is usually handled through the concept of memory
\cite{bennett1987,scully2007}.
In the case of the one-way barrier, the demon must somehow have a record of
the atomic positions.
If the demon knows the exact atomic positions, then to him each atom
occupies a very tiny volume in phase space.
By allowing the atoms into one side of the chamber, he is not really
compressing phase space, but rather rearranging it.
Rearranging phase space is possible via normal Hamiltonian evolution and is
not a violation of the second law of thermodynamics.

If the demon does not already know the positions, the demon must somehow
monitor the system and determine when to allow atoms through.
A measurement must take place, and some device must take action on the
result.
Therefore, the measurement result must be somehow imprinted in the device.
For the next measurement, either the result must be recorded elsewhere, or
the first result must be erased.
If the result is erased, that is an irreversible process which involves
dumping entropy into the environment.
If the result is not erased, then there is a memory record which goes from
a known initial state to a disordered state, containing all the disorder
that was removed from the atoms.
Thus, entropy for the combined device system does not decrease.

As a quick example, assume we have $N$ addressable atoms distributed
between two equal volumes $V_1$ and $V_2$.
A demon is preparing to make sure all the atoms are in $V_2$, but the
location of each atom within the two volumes is completely unknown.
Say the demon measures the position of each atom one by one, and if they
are found in $V_1$, then they are moved to $V_2$.
Once it is known in which volume an atom resides, a process that swaps
containers need not involve an entropy change, as the two volumes are of
equal size.
The demon now has a record of which volume each atom was found in.
In the simplest case, this memory is a series of $N$ bits each set to $1$
or $2$.
We will assume this memory to be constrained by the second law of
thermodynamics.
We will also assume that the initial state of the memory is known, since
writing a bit over an unknown state is the same as erasing, which, as
described below, requires dumping entropy into the environment.

If the demon erases the bit after each measurement, possibly to reuse it,
then after each measurement the demon must reduce the phase-space volume of
that bit by a factor of $2$.
This is because the erasure process must take \emph{two} states
(representing which volume contained the particle) and map them into
\emph{one} (the erased state).
Attempting to avoid this by measuring the state of the bit to decide how to
erase it just transfers the problem to some other memory that needs to be
erased.
By the second law of thermodynamics, this entropy decrease of
$k_\mathrm{B}\ln 2$ must be accompanied by a matching increase of entropy
elsewhere, so that the total entropy does not decrease.
After measuring and placing all $N$ atoms, the demon must have dumped
$Nk_\mathrm{B}\ln 2$ of entropy to the outside environment.

If the demon uses $N$ different bits and does not erase them, then after
the cycle is complete the memory is in one of $2^N$ equally likely states,
since each atom was in either $V_1$ or $V_2$ with equal probability.
To an outside observer the atoms are now in a smaller volume, but the
state of the demon's memory has gone from a known initial state (zero
entropy) to one of $2^N$ possible states, which is the same entropy
increase as if the demon had erased the bits and dumped that entropy into
the environment.
This also exactly matches the entropy \emph{decrease} of the atoms being
compressed from two identical volumes into one.
The observer cannot know the initial atomic distribution without having
made measurements of his own, either of the atomic positions or the demon's
memory.
In each case there is an entropy transfer.

Here is a short computer analogy.
Assume the atoms are in a one-dimensional trap, with the position
represented by a number stored in memory.
The volume is proportional to the number of states the number can represent,
and the entropy is proportional to the logarithm of that, which is
proportional to the number of digits, or bits, in that number.
The first bit of the number could represent whether the atom was located in
$V_1$ or $V_2$.
The demon above could be thought of as taking atoms from $V_1$ and placing
them into the exact same place within $V_2$.
In this analogy, that means forcing the first bit of the number to $1$,
effectively decreasing the size of the number by one bit.
The number of total digits (entropy) did not decrease; rather, one was
split off and transferred to the demon's memory, and no longer represents
part of the atom's position.

This demon-powered barrier is similar to our one-way barrier.
We can ignore extraneous effects such as photon recoil (which results in
heating), unwanted scattering, and beam jitter, and still account for
entropy.
All that is needed is the irreversible step and the corresponding
measurement record (memory) associated with it.
The ``memory'' for our one-way barrier is the repumping barrier beam, and
it accounts for the entropy even if we idealize it.
The ideal repumping barrier beam changes the state of atoms from $F{=}1$ to
$F{=}2$ and leaves $F{=}2$ atoms alone.
That irreversible step can take place because the transitions available to
the $F{=}1$ ground state are of a different frequency than those from those
available to the $F{=}2$ ground state.
When an atom is changed from the $F{=}1$ state to the $F{=}2$ state, a
repumping-barrier-beam photon is absorbed and a \emph{different} photon is
emitted.
Repump photons cannot change the atom from $F{=}2$ back to $F{=}1$, but the
lower-frequency emitted photons can.
Completely ignoring random spontaneous emission directions, we look only at
whether the frequency of each photon is resonant with the $F{=}1 \to F'$
transitions or not.
Each photon that is not resonant means that the number of trapped atoms was
increased by one.

Let the trapping volume occupy a fraction $r$ of the total volume $V$.
It can be shown that the optimum setup is to have the largest optical depth
possible, and as many untrapped atoms in the trapping region as possible.
The number of untrapped atoms would normally be proportional to the volume
ratio $r$ of the trapping region, but could be lower if many atoms had just
been trapped.
In this limit each repumping-barrier-beam photon increases the number of
trapped atoms with probability
\begin{equation}
\label{eq:P_trap}
\mathcal{P}_\mathrm{trap}
  = \frac{rN_1}{rN_1+N_2}
,
\end{equation}
where $N_1$ is the number of untrapped atoms, $rN_1$ is the number of
untrapped atoms in the trapping region, and $N_2$ is the number of trapped
atoms.

Say we trap some fraction $f$ of the $N$ atoms, and it takes $P$ photons to
do so.
We can use the entropy of a two-part ideal gas to write out the change in
entropy of the atoms (two-part because the trapped and untrapped atoms must
be distinguishable if the barrier is to tell them apart).
The atoms change from $N$ atoms in a volume $V$ to $\left(1-f\right)N$
untrapped atoms in a volume $V$ and $fN$ trapped atoms in a volume $rV$.
The change in entropy for the atoms is
\begin{equation}
\label{eq:atom_dS}
\Delta S_\mathrm{a} = k_{\scriptscriptstyle{\mathrm{B}}}N\left[
                          f\ln r
                        - \left(1-f\right)\ln\left(1-f\right)
                        - f\ln f
                      \right]
.
\end{equation}
Likewise, the repumping photons change from $P$ identical photons to $fN$
photons in one state and $P-fN$ photons in a different state, spread
throughout the same volume.
However, that change is more likely to occur toward the beginning, when
there are fewer trapped atoms that can become untrapped
[$\mathcal{P}_\mathrm{trap}$ is larger when $N_2$ is smaller in
Eq.~(\ref{eq:P_trap})].
We can approximate the entropy increase for the repumping beam by summing
up the entropy increase of the beam for each atom trapped.
Each term in the sum is the expected number of photons it will take to trap
the next atom times the entropy of a single photon in a mixed state, which
reflects the uncertainty of whether or not the photon was emitted by a
newly trapped atom.
The change in entropy of the repumping beam can thus be written as
\begin{eqnarray}
\nonumber
\Delta S_\mathrm{r} &=&
	-\sum_{n=0}^{fN-1} \frac{1}{\mathcal{P}_n} \left[
		  \mathcal{P}_n\ln \mathcal{P}_n
		+ \left(1-\mathcal{P}_n\right)\ln\left(1-\mathcal{P}_n\right)
	\right]
\\
\label{eq:photon_dS}
\mathcal{P}_n &=&
	\frac{rn}{rn+\left(N-n\right)}
,
\end{eqnarray}
where $\mathcal{P}_n$ is the probability of trapping another atom after $n$
atoms have been trapped, as given by Eq.~(\ref{eq:P_trap}).

Equation (\ref{eq:atom_dS}) shows a decrease in entropy from the decrease
in volume, but an increase in entropy due to the multiple atom types.
The increase disappears in the limit of all atoms being trapped ($f \to 1$).
The repumping barrier beam shows a net increase in entropy due to the
change in states of the photons.
This repumping-barrier-beam entropy increase can be shown (at least in a
large-number approximation) to always be at least as large as the decrease
in atom entropy, so the net entropy change is non-negative.
Thus, the changes in photon frequency are sufficient to uphold the second
law of thermodynamics.
This analysis does not include momentum diffusion for either atoms or
photons resulting from photon recoil, so entropy actually increases more
than is shown here.

We created a version of such a demon.
Figure \ref{fig:compression} shows data from an experiment where we had
atoms spread throughout a volume, and then activated a one-way barrier in
the center.
The transfer was not completely efficient---not all atoms were trapped on
the right-hand side of the barrier, but the spatial compression is quite
good.
However, the spatial compression is almost completely countered by
heating from spontaneous scattering of the barrier-beam light.
We found that the overall phase-space volume occupied by atoms during this
experiment decreased by $7(2)\%$.
Better compression is likely achievable should we optimize the setup for
phase-space compression; presently we optimize the barrier for transmission
rather than compression.
To achieve better phase-space compression, we would use a smaller, slowly
moving barrier starting initially off to one side of the atom distribution,
as described by Ruschhaupt \textit{et~al.}\ \cite{ruschhaupt2006c}.
At first, this barrier would capture only the atoms with the highest
energies, and would do so near their turning points so that they had very
little kinetic energy (allowing for a weaker barrier and less scattering).
We would then adiabatically sweep the barrier along the dipole-trap axis,
through the focus of the dipole trap.
At each point, the barrier would trap new atoms near their turning points,
where they had very little kinetic energy.
The atoms that were already trapped would not gain kinetic energy.
Thus when the barrier passed through the dipole-trap focus, all the atoms
would be left in the center of the trap, with less kinetic energy.

\section{\label{sec:closing}Closing}

In summary, we have demonstrated an all-optical asymmetric potential
barrier capable of increasing the overall phase-space density of a sample
of neutral alkali atoms.
Furthermore, we have found that the barrier is robust, with fairly
substantial changes having little effect, including mechanical changes,
optical changes, and changes in atom-state preparation.
We have also developed some theory and simulations describing the barrier.

The authors would like to thank Eryn Cook and Paul Martin for their helpful
comments.
This research was supported by the National Science Foundation, under
Project No.\ PHY-0547926.

\appendix

\section{\label{app:state_change}State Changing Probabilities}

Here we review the computation of approximate probabilities for linearly
polarized light to change the state of an atom, for use in our simulations.
In the dipole approximation, an atom must be excited by the incident light
field and then spontaneously decay to a different ground state in order for
the state to change.
We also assume that the incident light field is a single monochromatic
beam.
Therefore, we wish to compute the amplitude for an atom changing from one
state ($F_0,m_0$) to another ($F_1,m_1$), via an excited state ($F',m'$),
and then sum over the excited states and square that sum to obtain a
probability for changing states.
We make the approximation that the $\mathrm{D}_2$ multiplet is dominant,
and further that the splitting between the lines is small compared to the
barrier-beam detuning.
With these approximations we find the probability to be
\begin{eqnarray}
	\label{eq:prob}
	\lefteqn{ P_{F_0,m_0 \to F_1,m_1} \propto }
	\\
	\nonumber
	&&
	\sum_{q'}\left|\sum_{F',m'}
	\left\langle F_1,m_1\right|\hat{d}_{q'}\left|F',m'\right\rangle
	\left\langle F',m'\right|\hat{d}_q\left|F_0,m_0\right\rangle
	\right|^2
,
\end{eqnarray}
where $q$ represents the polarization of the incident light, $q'$
represents the polarization of the spontaneously emitted light ($-1$, $0$,
and $+1$ for the $z$ component of angular momentum), and $\hat{d}_q$ and
$\hat{d}_{q'}$ are the appropriate spherical-basis dipole operators.
The polarization of the incident laser field determines $q$, but we sum
over all $q'$ and all intermediate states ($F',m'$) within the subset of
excited states that are close to the laser frequency.
The $q'$ sum is done after squaring the amplitude because different emitted
polarizations are distinguishable.
The ($F',m'$) sum is performed with the amplitudes because the
near-degeneracy assumption means the different emissions are
indistinguishable.
Since this does not represent \emph{all} possible excited states (or any
ground states), the sum of $\left|F'm'\right\rangle\left\langle
F'm'\right|$ is not the identity.

Next, we average over the initial magnetic sublevel $m_0$ and the emission
polarization $q'$ because we do not measure or simulate the magnetic
sublevels of the atoms.
This is also an approximation, and is partially justified by the fact that
the barrier operation was not significantly affected by the incident light
polarization.

Finally, we can conserve angular momentum by requiring $m'+q'=m_1$ and
$m_0+q=m'$.
We can then evaluate Eq.~(\ref{eq:prob}) and perform the averages over
the magnetic sublevels of $^{87}\mathrm{Rb}$.
Given that we use linearly polarized light ($q=0$) in our experiment, we
find
\begin{equation}
	\label{eq:bar_F_state}
	\begin{array}{rclcl}
		\displaystyle
		P_{\mathrm{barrier,}F{=}1 \to F{=}2} &=& 5/18 &=& 0.2\bar{7}
		\\
		\displaystyle
		P_{\mathrm{barrier,}F{=}2 \to F{=}1} &=& 1/6 &=& 0.1\bar{6}
		.
	\end{array}
\end{equation}
These are the values we use for the probability of a barrier-beam
scattering event changing the state of an atom.
The repumping barrier light is tuned close to the $F{=}1 \to F'{=}2$
transition, so we require $F'{=}2$ in the $F'$ sum, with the following
result:
\begin{equation}
	\label{eq:rep_F_state}
	\begin{array}{rclcl}
		\displaystyle
		P_{\mathrm{repump,}F{=}1 \to F{=}2} &=& 1/2 &=& 0.5
		.
	\end{array}
\end{equation}
We found that the simulations were not particularly sensitive to these
probabilities, although the barrier performed better in simulations with
these ratios than in simulations where all ratios were $50\%$.
As a side note, we also computed the probabilities that took into account
the different detunings of the excited states by weighting the excitation
amplitudes in Eq.~(\ref{eq:prob}) by factors inversely proportional to the
detunings.
The probabilities then became
\begin{equation}
	\label{eq:bar_F1_detunings}
	\begin{array}{rclcl}
		\displaystyle
		P_{\mathrm{barrier,}F{=}1 \to F{=}2} &\approx& 0.271
		\\
		\displaystyle
		P_{\mathrm{barrier,}F{=}2 \to F{=}1} &\approx& 0.114
		\\
		\displaystyle
		P_{\mathrm{repump,}F{=}1 \to F{=}2} &\approx& 0.500
		.
	\end{array}
\end{equation}
These changes were not enough to significantly alter the simulations.

\section{\label{app:analytic}Analytic scattering solution}

\begin{figure}[tbp]
	\begin{center}
		\includegraphics[width=\columnwidth]{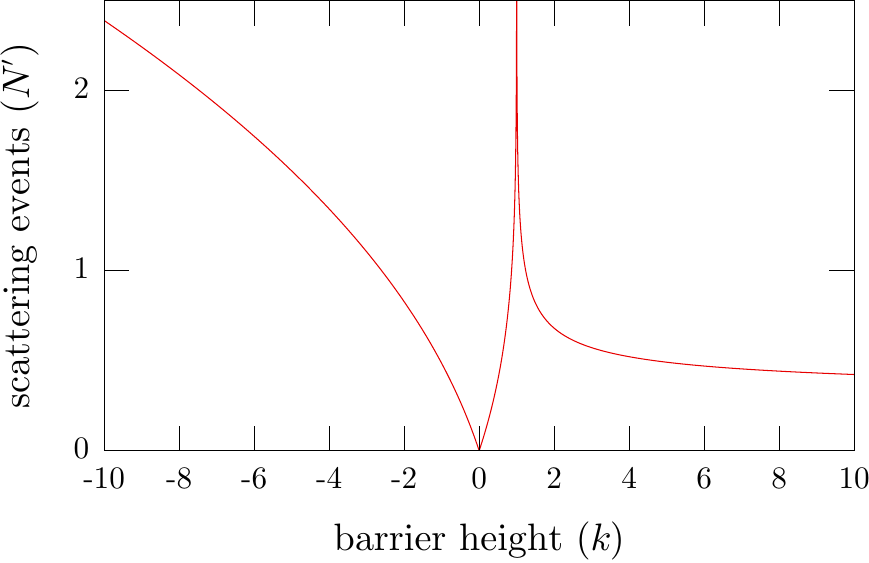}
	\end{center}
	\vspace{-5mm}
	\caption{%
		(Color online)
		Plot of Eq.~(\ref{eq:N_scatter}), showing the expected number
		of scattering events for a two-level atom passing through a
		Gaussian barrier.
		The barrier height $k$ is defined in Eq.~(\ref{eq:scatter_k}), and the
		number of scattering events is given as
		$N':=Nw_0mv_0\Gamma/\hbar\left|\Delta\right|$, where $N$ is the actual
		number of scattering events.
	\label{fig:scatter_k_plot}}
\end{figure}

\begin{figure}[tbp]
	\begin{center}
		\includegraphics[width=\columnwidth]{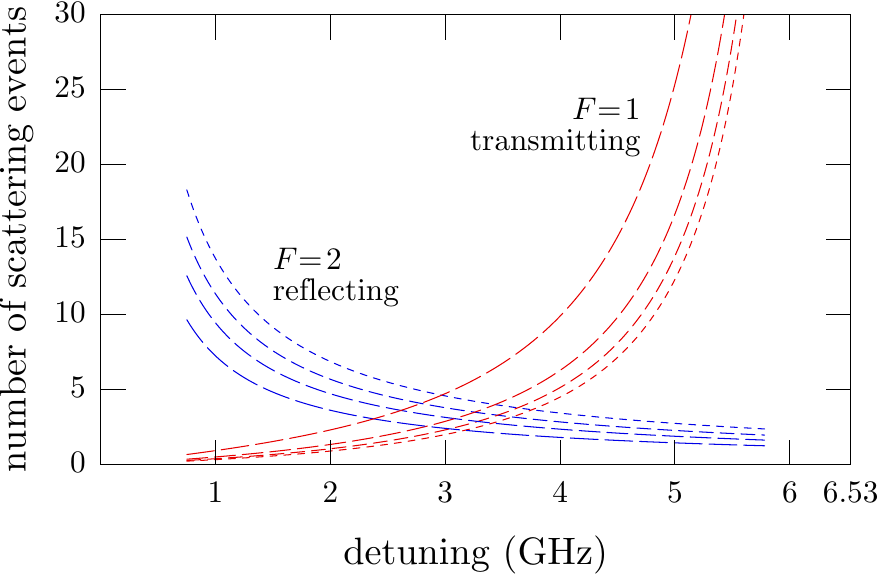}
	\end{center}
	\vspace{-5mm}
	\caption{%
		(Color online)
		Plot of Eq.~(\ref{eq:N_scatter}), showing the expected number of
		scattering events for a quasi-two-level (ignoring state changes)
		$^{87}\mathrm{Rb}$ atom passing through a Gaussian barrier as a
		function of beam detuning from the $F{=}2 \to F'$ transitions.
		The curves that increase for larger detunings are for atoms in the
		$F{=}1$ transmitting state, and the curves that decrease for larger
		detunings are for atoms in the $F{=}2$ reflecting state.
		The effective detuning from the $F{=}1$ ground state for a beam tuned
		to the $F{=}2\to F'$ transitions is about $6.53 \ \mathrm{GHz}$.
		Each $F{=}1$ and $F{=}2$ pair of curves represents a different barrier
		height.
		The dotted curves show the expected numbers of scattering events for a
		barrier height of $k=1.25$.
		This means that the beam intensity for each detuning is adjusted so
		that the potential barrier seen by atoms in the $F{=}2$ ground state is
		$1.25$ times the kinetic energy of the atoms.
		The short-dashed curves are for $k=1.5$, the medium-dashed curves are
		for $k=2$, and the long-dashed curves are for $k=4$.
		Our canonical data had $k=2.3$ with an effective two-level detuning
		of $1.12 \ \mathrm{GHz}$.
	\label{fig:scatter_d_plot}}
\end{figure}

We can compute how many photons, on average, a two-level atom scatters
while crossing a Gaussian beam if we ignore momentum kicks due to
spontaneous emission.
The barrier width is small compared to the Rayleigh length of the
dipole trap, so we ignore any changes in the underlying potential and treat
an atom in free space crossing a Gaussian potential with a scattering rate
proportional to that potential.
For a two-level atom in the far-detuned, rotating-wave approximation, the
scattering rate is $R=\left|\Gamma V/\hbar\Delta\right|$, where $\Gamma$ is
the transition linewidth, $V$ is the ac Stark shift (optical potential),
and $\Delta$ is the (angular) detuning.
The average number of scattering events can be computed by integrating the
local scattering rate for all time, which gives
\begin{equation}
\label{eq:N1}
N = \int_{-\infty}^\infty R\left[x{\left(t\right)}\right] \ dt
,
\end{equation}
where $x(t)$ is the location of the atom at time $t$.
For a Gaussian potential, we know the atom will approach the barrier and
then either reflect or transmit.
The expected number of scattering events up until the reflection (or
passage through the barrier center) is the same as after the event, so we
can restrict the integral to a period where the velocity does not change
sign.
This allows the change of variables to $x$, using $dt = dx / v(x)$, where
$v(x)$ is the velocity of the atom when it is at position $x$.
After this change of variables, we find 
\begin{equation}
\label{eq:N2}
N = 2\int_{x_0}^\infty \frac{R{\left(x\right)}}{v{\left(x\right)}} \ dx
,
\end{equation}
where $x_0=0$ in the case where the atom passes through the center of
the barrier, and the turnaround point in the case where it does not.

If we let $w_0$ be the $1/e^2$ intensity radius of the Gaussian beam, then
we can use conservation of energy to compute
\newlength{\colonequal}
\settowidth{\colonequal}{$:=$}
\begin{eqnarray}
\label{eq:vofx}
v{\left(x\right)} &\makebox[\colonequal][r]{=}& v_0\sqrt{1-ke^{-2x^2/w_0^2}}
\\
\label{eq:scatter_k}
k &\makebox[\colonequal][r]{:=}& \frac{V_0}{\frac{1}{2}mv_0^2}
,
\end{eqnarray}
where $v_0$ is the initial velocity of the atom [$v_0 = v(x\to\infty)$],
$V_0$ is the height of the barrier, and $k$ is defined as the ratio of the
barrier height to the incident kinetic energy of the atom.
We can compute $x_0$ by solving $v(x)=0$:
\begin{equation}
\label{eq:x0}
x_0 = \left\{
	\begin{array}{ll}
		\displaystyle
		0 & k\le 1
		\\
		w_0\sqrt{\frac{\ln k}{2}} & k>1
		\displaystyle
	\end{array}
\right.
.
\end{equation}
A negative value for $k$ means the barrier is actually an attractive well,
and $k < 1$ means the barrier is not high enough to stop the atom.
For these cases, $x_0=0$.
Substituting the scattering rate $R(x)$ and the velocity $v(x)$ from
Eq.~(\ref{eq:vofx}) into Eq.~(\ref{eq:N2}) gives
\begin{equation}
\label{eq:N3}
N = \frac{mv_0\Gamma\left|k\right|}{\hbar\left|\Delta\right|}
	\int_{x_0}^\infty \frac{e^{-2x^2/w_0^2}}{\sqrt{1-k e^{-2x^2/w_0^2}}} \ dx
,
\end{equation}
where $x_0$ is given by Eq.~(\ref{eq:x0}).

We now change the integration variable in Eq.~(\ref{eq:N3}) to
$u=\exp(-2x^2/w_0^2)$, with the result
\begin{equation}
\label{eq:N_scatter}
N =
\frac{w_0mv_0\Gamma\left|k\right|}{2\sqrt{2}\hbar\left|\Delta\right|}
	\int_{0}^{\left|\max{\left\{1,k\right\}}\right|^{-1}}
	\frac{du}{\sqrt{1-ku}\sqrt{\ln\left(\frac{1}{u}\right)}}
.
\end{equation}
The $k$ limits are as expected:
\begin{enumerate}
	\item
	$N\to 0$ as $k\to 0$.
	The number of scattering events, $N$, converges linearly to zero as the
	barrier height $k$ (or well depth, if $k$ is negative) decreases.
	When there is no barrier, there is no scattering.
	\item
	$N\to\infty$ as $k\to 1$.
	This is the usual logarithmic divergence (it diverges as
	$-\ln\left|1-k\right|$) of the time it takes an atom to roll off a
	hilltop starting with no initial velocity.
	In this limit, the number of scattering events, $N$, diverges because the
	barrier height $k$ is approaching the limit where the atom comes to rest
	at the top of the barrier (and stays there for an infinite amount of
	time).
	\item
	$N\to\infty$ as $k\to -\infty$.
	The number of scattering events, $N$, diverges as $\sqrt{|k|}$ as the
	potential well depth $|k|$ becomes infinite, meaning that for a very deep
	attractive well the increase in speed with which the atom passes through
	the barrier is not enough to counter the increase in scattering rate.
	Interestingly, if we keep the well depth constant and let $k\to -\infty$
	by decreasing the initial kinetic energy (let $v_0\propto 1/\sqrt{|k|}$),
	$N$ converges to a constant.
	This is because in this limit, the initial speed of the atom is
	negligible compared to the speed increase as it crosses the attractive
	well, so the initial speed, and thus $k$, does not matter.
	\item
	$N\to 0$ as $k\to\infty$.
	The number of scattering events $N$ converges as $1/\sqrt{\ln k}$ as the
	barrier height $k$ becomes infinite, and means that if the barrier is
	much higher than it needs to be to reflect an atom, the number of
	scattering events can be decreased to zero.
\end{enumerate}

Equation (\ref{eq:N_scatter}) shows why changing how far back we initially
start the atoms is not quite the same as changing the barrier height.
Changing the barrier height affects only $k$, whereas changing the initial
location also affects $v_0$.

Figure \ref{fig:scatter_k_plot} shows a plot of Eq.~(\ref{eq:N_scatter}).
Note the weak dependence on barrier height for a reflecting barrier of at
least twice the required height ($k>2$).
Also note the relatively strong dependence of scattering events on well
depth for attractive wells ($k<0$).
This hints at another reason why we want the main barrier beam tuned closer
to the $F{=}2$ ground-state transitions than the $F{=}1$ ground-state
transitions.
The main barrier beam will not scatter much on reflection if it presents a
reasonably high barrier, even if the detuning is not very large.
However, the transmitting potential well that the main barrier beam
presents to the $F{=}1$ ground-state atoms quite possibly will scatter many
photons, because the scattering rate increases much faster with $|k|$ for
an attractive well.
This can be seen in Fig.~\ref{fig:scatter_d_plot}, where we plot, for a few
different barrier heights, the expected number of scattering events for
each state of the atom as a function of beam detuning.
The main-barrier-beam intensity was constrained to maintain a constant
(independent of detuning) barrier height as seen by $F{=}2$ atoms.
Since atoms only need to transmit once but must reflect many times, one
might think from Fig.~\ref{fig:scatter_d_plot} that the optimum barrier
frequency is around $4 \ \mathrm{GHz}$ from the $F{=}2 \to F'$ transitions.
However, as discussed earlier, scattering events for transmitting atoms are
much worse than for reflecting atoms because they may change the state of
the atom to the reflecting state, causing much more heating and loss than
the reverse situation.
We found that a detuning that reduced scattering events on transmission
much more than on reflection was optimum.
For the experiments we performed (with the exception of experiments where
we changed how far back the atoms started), $k$ was about $2.3$ for atoms
in the reflecting state.
For atoms in the transmitting state (with the exception of experiments
where either the frequency of the barrier or the initial position was
varied), $k$ was about $-0.47$.
With our effective two-level detuning value of $1.12 \ \mathrm{GHz}$ for
the reflecting barrier, this yields $\sim\! 8$ scattering events on
reflection, and $\sim\! 0.7$ scattering events on transmission (with an
effective two-level detuning of $-5.41 \ \mathrm{GHz}$).

\section{\label{app:ang_mom}Angular momentum in single-focused-beam dipole traps}

\begin{figure}[tbp]
	\begin{center}
		\includegraphics[width=\columnwidth]{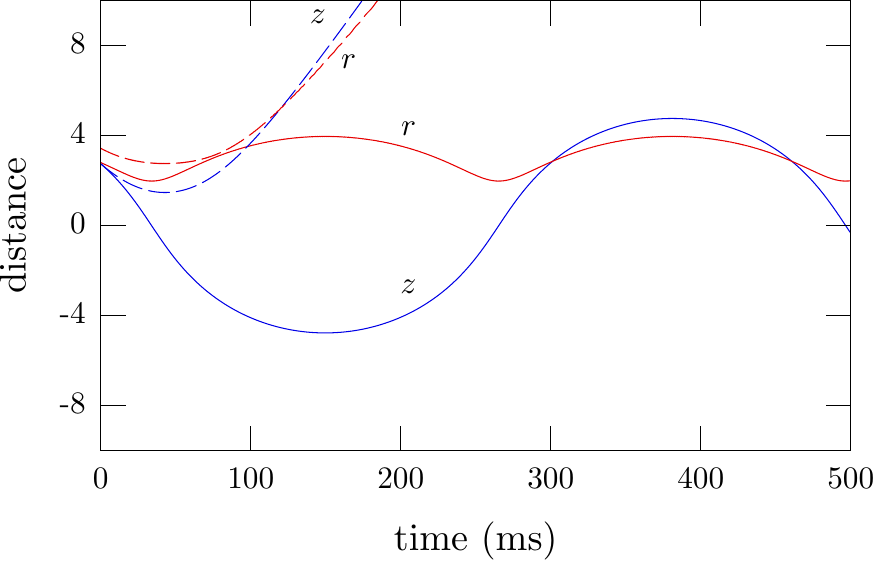}
	\end{center}
	\vspace{-5mm}
	\caption{%
		(Color online)
		Plot of cylindrical coordinates for two orbits near the focus of a
		dipole trap using the same dipole-trap parameters as in
		Sec.~\ref{sec:simulation}, except without gravity.
		Longitudinal distances ($z$) are in millimeters, and radial distances
		($r$) are in ten-micron units.
		The solid curves show a simulation with $90\%$ of the critical angular
		momentum required for the dipole-trap focus to become repelling.
		This trajectory shows an oscillation about the focus that is slowed due
		to the angular momentum.
		The dashed curves show the same initial conditions, but with a slightly
		larger $r$ value and azimuthal velocity, pushing the angular momentum
		to $110\%$ of the critical value, and show how such a particle can
		appear to bounce off an attractive potential.
	\label{fig:ang_focus}}
\end{figure}

\begin{figure}[tbp]
	\begin{center}
		\includegraphics[width=\columnwidth]{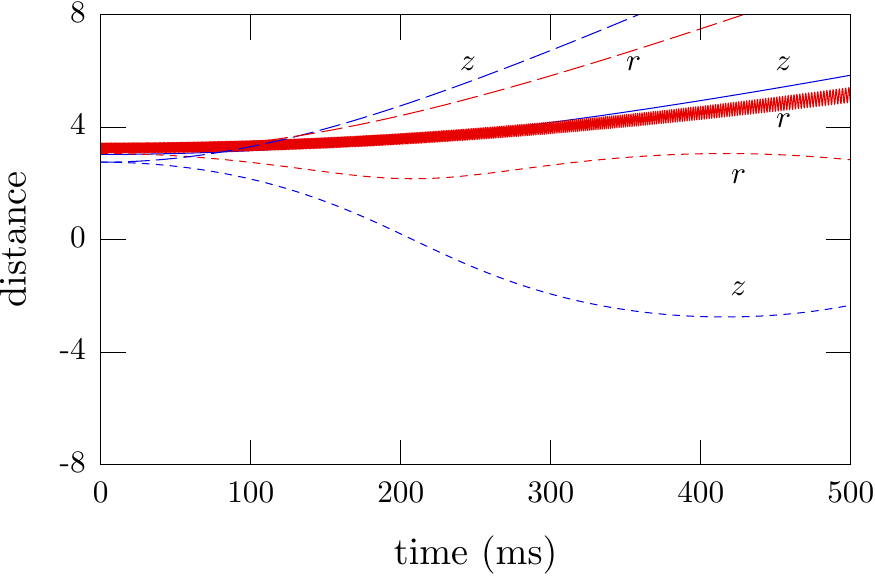}
	\end{center}
	\vspace{-5mm}
	\caption{%
		(Color online)
		Plots of cylindrical coordinates for three orbits in a dipole trap,
		using the same dipole-trap parameters as in Sec.~\ref{sec:simulation},
		except without gravity.
		Longitudinal distances ($z$) are in millimeters, and radial distances
		($r$) are in ten-micron units.
		The solid curve shows a slight variation of the predicted saddle-point
		orbit, with the correct angular momentum but with $q$ slightly below
		unity.
		This curve shows the particle oscillating about the saddle-point orbit,
		but drifting along the $q{=}1$ manifold.
		The radial restoring force for this orbit in our trap is much stiffer
		than the longitudinal restoring force, so $r$ oscillates rapidly
		(resulting in a noncircular orbit).
		The dashed curve shows the predicted saddle-point orbit, but with
		$101\%$ of the saddle-point angular momentum; this orbit diverges.
		The dotted curve shows the predicted saddle-point orbit, but with
		$99\%$ of the saddle-point angular momentum; this trajectory falls away
		from the orbit and instead oscillates about the focus.
	\label{fig:ang_not_focus}}
\end{figure}

In this appendix we briefly describe how angular momentum affects atomic
motion in a dipole trap formed by a single focused Gaussian beam.
The basic effect is that an atom with angular momentum about the trap axis
feels an effective force that repels it from the trap focus.
A similar effect is experienced by a ball rolling inside a funnel.
The tighter confinement toward the center of the funnel (or dipole-trap
focus) forces a higher angular velocity in order to conserve angular
momentum, which results in an outward push.
This effect is often dealt with in classical mechanics textbooks
such as Ref.~\cite{marion1988} when dealing with central potentials.
The angular momentum of a particle in a central potential leads to a
centrifugal potential-energy term that repels the particle from the center
of the potential.

For this appendix, we will assume we have only the focused Gaussian
dipole-trap beam, creating a conservative potential proportional to the
local intensity.
The potential can thus be written as
\begin{equation}
\label{eq:dipole_V}
V{\left(r,z\right)} =
	-V_0 \frac{\exp\left({\displaystyle -\frac{2r^2}{1+z^2}}\right)}{1+z^2}
,
\end{equation}
where $r$ is the radial coordinate measured from the trap axis, scaled by
the $1/e^2$ intensity radius at the focus, and $z$ is the longitudinal
coordinate measured from the trap focus, scaled by the Rayleigh length of
the beam.
We are discussing attractive traps, so $V_0$ is assumed to be positive.
The quantity in the exponential in Eq.~(\ref{eq:dipole_V}) will occur
several times in this appendix, so we will abbreviate it as $q^2$:
\begin{equation}
\label{eq:dipole_q}
q^2 := \frac{2r^2}{1+z^2}
.
\end{equation}
With this abbreviation, the equations of motion for the cylindrical
coordinates become
\begin{eqnarray}
\label{eq:dipole_z_acc}
\frac{d^2 z}{dt^2} &=&
	- \frac{2 V_0}{m z_0^2} z \left(1-q^2\right)
	\frac{\exp\left(-q^2\right)}{\left(1+z^2\right)^2}
\\
\label{eq:dipole_r_acc}
\frac{d^2 r}{dt^2} &=&
	\frac{1}{m^2 w_0^4 r^3}
	\left[J^2 - m w_0^2 V_0 q^4 \exp\left(-q^2\right) \right]
.
\end{eqnarray}
Here $J$ is the conserved angular momentum about the trap axis, and $z_0$
and $w_0$ are the Rayleigh length and $1/e^2$ intensity radius,
respectively.
Not surprisingly, the focus ($z{=}0$) is a fixed point of the longitudinal
motion.
What is more surprising is the $(1-q^2)$ factor.
The presence of this factor means that circular orbits with large angular
momenta can have significantly smaller accelerations along the trap axis
than for one-dimensional motion, even attaining a different sign.
We observed experimentally that our trap period was longer than what we
would expect from a purely one-dimensional treatment.
In simulations we could see that this was partially due to atoms in
high-angular-momentum orbits taking a longer time to reach the trap focus.
We could also observe the repulsion from the trap center in simulations.
This occurs when the energy required for a given angular momentum at the
focus is larger than the trap depth.
Figure \ref{fig:ang_focus} shows simulations of some trajectories with
these effects.
We took the same simulations used in Sec.~\ref{sec:simulation}, but with
single atoms and initial conditions appropriate for orbits (and no gravity,
in order to preserve cylindrical symmetry) near the critical point,
illustrating how the focus of the dipole trap changes from attracting to
repelling.

This effect helps explain the difference between
Figs.~\ref{fig:sim_left_right} and \ref{fig:sim_left_right_1D}.
Atoms with some angular momentum have $q>0$, reducing the longitudinal
acceleration toward the trap focus in Eq.~(\ref{eq:dipole_z_acc}).
This is why the simulation in Fig.~\ref{fig:sim_left_right_1D} shows atoms
crossing the trap center sooner than the simulation in
Fig.~\ref{fig:sim_left_right}.
Furthermore, because atoms without angular momentum experience more
longitudinal acceleration, they are traveling faster along the trap axis
when they cross the barrier than their counterparts with angular momentum.
We believe this is the main reason why the atoms traverse back and forth
across the barrier in Fig.~\ref{fig:sim_left_right_1D} more easily than in
Fig.~\ref{fig:sim_left_right}---the barrier is effectively higher for atoms
that are moving more slowly along the trap axis.

Linear stability analysis of Eqs.~(\ref{eq:dipole_z_acc}) and
(\ref{eq:dipole_r_acc}) suggests that for $q^2=1$, there exist
``saddle-point orbits'' at some point that is \emph{not} the trap focus.
These orbits require $q^2=1$ and $J^2=m w_0^2 V_0 \exp(-1)$, and are stable
to perturbations in $q$ but not to perturbations in $J$.
They have no linear restoring force along the $q^2{=}1$ manifold.
We were able to produce simulations that show these strange orbits, but
they were so sensitive that we had to remove the effects of gravity from
our simulations in order for them to persist.
Figure \ref{fig:ang_not_focus} shows one of these orbits and demonstrates
how, even with the right conditions, the orbit drifts along the $q^2{=}1$
curve.
This drift is a result of nonlinear terms in the equations, which dominate
because the linear term is zero.
Figure \ref{fig:ang_not_focus} also shows how quickly orbits with the wrong
angular momentum diverge.

\end{document}